\documentclass[onecolumn]{aastex6}

\newcommand{\beq}{\begin{equation}}
\newcommand{\eeq}{\end{equation}}
\newcommand{\myemail}{jiangyg@sdu.edu.cn}

\usepackage{amsmath}
\usepackage{natbib}
\usepackage{amsfonts,amssymb}
\usepackage{color}
\usepackage{graphicx}

\begin{document}


\title{Locations of optical and $\gamma$-ray emitting regions in the jet of PMN J2345-1555}

\shorttitle{Multiwave analysis of PMN J2345 for seven years data }
\shortauthors{Jiang et al.}

\author{Yun-Guo Jiang\altaffilmark{1}, Shao-Ming Hu\altaffilmark{1}, Xu Chen\altaffilmark{1}, Xi Shao\altaffilmark{1},Qiu-Hong Huo\altaffilmark{2}}
\affil{\altaffilmark{1}Shandong Provincial Key Laboratory of Optical Astronomy and Solar-Terrestrial Environment,  \\ Institute of Space Sciences, Shandong University, Weihai, 264209, China ; \myemail}

\affil{\altaffilmark{2} School of Space Science and Physics, Shandong University at Weihai,  Weihai 264209, China}

\begin{abstract}
We collect long term  $\gamma$-ray, optical and radio $15$ GHz  light curves of  quasar  object PMN J2345-1555. The correlation analyses between them are performed via the  local cross-correlation function (LCCF). We found that all the optical $V$, $R$ band and the infrared $J$ band are  correlated with the radio 15 GHz at beyond  $3\sigma$ significance level, and the lag times are $-221.81^{+6.26}_{-6.72}$,  $-201.38^{+6.42}_{-6.02}$ and $-192.27^{+8.26}_{-7.37}$ days, respectively. The $\gamma$-ray is strongly correlated with optical, but weakly correlated with the radio.   We present that time lags between different frequencies can be used as an alternative parameter to derive the core-shift measurement. For this target, the magnetic field and particle density at 1 parsec in jet are derived to be $0.61$ Gauss and $1533/\gamma_{\rm min}$ cm$^{-3}$, respectively. The black hole mass and the 15 GHz core position in jet are estimated to be $10^{8.44} {\rm M}_{\odot}$ and $30$ parsec, respectively. The lag times enable us to derive that the optical  and the $\gamma$-ray  emitting regions  coincide, which are located at $4.26^{+0.83}_{-0.79}$ pc away from 15 GHz core position in jet and beyond the broad line region (BLR).
We found that a $3\sigma$ correlation between the color index and the radio light curve, which indicates that opacity may play an important role in the variation. The $\delta V-\delta R$ behaviors are complex, while the $R-J$ shows a bluer when brighter trend. As hinted from radio images, we proposed a positional dependent spectral index model to explain the color index behaviors, which is complementary for the shock in jet model. The  curvature effects  and contribution from accretion disk may also affect variables of blazars in many aspects.
\end{abstract}




\section{Introduction} \label{sec:intro}

In the unified model of active galactic nuclei (AGNs), blazars are of  a special subclass whose relativistic jets point towards us, characterised by its violent variable in the whole electro-magnetic bands.
The spectral energy distribution (SED) of blazars indicates two bumps, which are interpreted as the synchrotron and Compton processes, respectively. These emissions are believed to originate from jets, whose axial direction
is near our line of sight. Except several nearby sources, blazars are point sources for optical telescopes and $\gamma$-ray detectors. Thus, the locations of the optical and $\gamma$  regions  are unsolved by observation directly.
The multi-frequency data help to answer this question from different aspects. Some indirect methods have been proposed to study the location of $\gamma$-ray emission regions based on the rich data from {\it Fermi} large area telescope (LAT). By SED fitting with the one zone emission model,  Kang et al. found that the best fitted parameters are obtained when the seed photons are from dust torus, which indicates that the $\gamma$-ray emission region is beyond the broad line region (BLR) \cite{Kang:2014}.
Britto et al.  studied one $\gamma$-ray outburst of 3C 454.3, and pointed out that gamma-ray emission region is on the edge of the BLR based on the minimal variable time scale \cite{Britto:2016}. If this time scale is considered as the lower limit of the cooling time scale, the magnetic field can be derived via the synchrotron radiation process \cite{Yan:2018}. Respecting the scaling law, the location of the emission regions can be derived by  matching the obtained value $B‘_{\rm diss}$ with  results from the Very Long Baseline interferometry (VLBI) core-shift measurements. Using the similar method, Wu et al. first obtained the strength of the magnetic field by SED fitting, then derived locations of $\gamma$-ray emission region by the scaling law \cite{Wu:2018}. However, the SED fitting process depends on free parameters, the coupling of them brings us a caution that best fitted results may deviate from the true physical values.

Impressive progresses have been made by several radio observation programs of AGNs in recent years \cite{Richard:2011,Pushkarev:2012,Fuhrmann:2016,Lister:2009,Lister:2019}. The high spatial resolution (milliarcsecond) of VLBI enables us to measure  parameters of jets directly.  The combination of VLBI radio images and the X-ray $\gamma$-ray light curves can provide a direct method to locate the emission regions.
Hodgson et al.  made long term VLBI observations for OJ 287, $\gamma$-ray data analysis combined with the radio images reveals that the $\gamma$-rays are radiated from the radio core regions, and strongly correlates with the jet structures in radio images \cite{Hodgson:2017}. Also, Algaba et al. found that outbursts of $\gamma$-ray of 4C 38.41 can be traced back with the appearance of new knot structures in radio images \cite{Algaba:2018}. It is evident  that gamma-ray originates from down stream regions in the jet.  The discrete correlation fuction (DCF) analysis indicates that $\gamma$-ray leads to radio about $70$ days, which corresponds to a 39 parsec distance between them.
Time lags can be used in a direct method to locate the $\gamma$-ray emission regions \cite{Fuhrmann:2014,Max:2014b}. Using the discrete cross-correlation function (DCF) stacking technique, Fuhrmann et al. found significant correlations between $\gamma$-ray and multiple radio wavelengths in a sample with 54 blazars \cite{Fuhrmann:2014}. The $\gamma$-ray emitting region is located  upstream of 3 mm radio core regions. \citet{Max:2014b} studied the correlation between $\gamma$-ray and  15 GHz radio light curves by using the local cross-correlation function (LCCF). They found that four blazars in a sample with 41 sources have strong correlations with significance level beyond $2.25 \sigma$. Compared to DCF, LCCF performs better in picking out the physical correlations \cite{Max:2014a}. The location of the $\gamma$-ray emitting region can be calculated by the time lag analysis \cite{Fuhrmann:2014,Max:2014b}. Combined with jet parameters from VLBI results, this provides a direct method to derive the emitting regions.

PMN J2345-1555 (B1950 name 2342-161) was found  in the Parkes-MIT-NRAO (PMN) 4850 MHz tropical surveys \cite{Griffith:1994}. Its position is localized  in the fourth VLBA calibration survey (VCS4), and the J2000 coordinates are R.A.: 23:45:12.4 and Decl.: -15:55:08 \cite{Petrov:2006}. Its radio brightness at 4.8 GHz is larger than 65 mJy and it has a flat spectrum of radio \cite{Healey:2007}.  Its redshift is 0.621, given by Healey et al in the optical characterization for a sample of bright blazars (CGRaBs) \cite{Healey:2008}. The optical spectroscopy identification is also performed by Shaw et al. with 5 m Hale telescope at Palomar \cite{Shaw:2009}. The optical continuum flux and the spectral index are $F_{\nu,10^{14.7}}=0.31$ mJy and $\alpha=1.3\pm0.02$ ($F_{\nu} \equiv \nu^{-\alpha}$), respectively.  This source is variable at all broad band frequencies. In January 2010, a $\gamma$-ray flare has been detected by Fermi large-area-telescope (LAT) \cite{Ciprini:2010}. In October 2010, a quasi-simultaneous flare has been detected in optical, near-infrared (NIR), X-ray and $\gamma$-ray \cite{Donato:2010,Carrasco:2010, Kaufmann:2010,Mathew:2010}. Carrasco et al. detected two  NIR flares in 2012 and 2013 respectively, and   more than 1.6 magnitude change in $H$ band has been reported \cite{Carrasco:2012,Carrasco:2013}.  This target has also been monitored by Steward observatory, SMARTS and KAIT in a long term manner \cite{Li:2003,Smith:2009,Bonning:2012}. Using the one-zone lepton emission model, Ghisellini et al. studied the SED at low and high luminosity epoches, and found that the target turns from a FSRQ to a BL Lac object. This behavior is interpreted as the emission zone moves from the BLR region to just outside\cite{Ghisellini:2013}.



 In the paper, we will study the location of optical and gamma-ray emission regions via the time lag analysis, which will provide a preliminary test for the one-zone emission models. More than 8 years data in $\gamma$-ray, optical and radio bands are collected in Section 2.  The details of LCCF and time lag analysis are presented in Section 3. We found significant correlations between optical $V$, $R$ and radio light curves. The time resolved color index analysis reveals that this source becomes bluer when brighter. In Section 4, the theoretical frame is reviewed first. Then, we derive the location of emitting regions for optical and $\gamma$-rays for this target. The magnetic field strength and number density of radiative particles are derived according to the scaling law, which is a direct method to derive these jet parameters. The strong correlation between color index and the light curve gives us the hint that the optical emission region has a spectral distribution structure similar to that of radio. We propose that this provides a new mechanism to explain the variation of color index behaviors.

\section{Data collection and reduction}
We collect the multiple wavelength data of the target from publicly available webs, including the Fermi LAT gamma-ray data \cite{Atwood:2009}, the optical photometry and polarization from Steward Observatory \cite{Smith:2009}, the 15 GHz radio data from Owens Valley Radio Observatory (OVRO) \cite{Richard:2011}.

{\bf Fermi LAT data} More than nine years (from 2008 August 4 to 2017 September 6) $\gamma-$ray data  of PMN J$2345-1555$ were retrieved from the public {\it Fermi Science Support Center} (FSSC). The downloaded data field is inside a region of interest (ROI) with $15^{\circ}$ radius around the target, and the photon energy interval is  $0.1 \sim 300$ GeV.
The data analysis was performed by using the {\it Fermi ScienceTools} version v10r0p5. The instrument response function is the ${\rm P8R2\_SOURCE\_V6}$.  The Galactic diffuse and isotropic $\gamma-$ray background are counted by the ${\rm gll\_iem\_v06}$ and ${\rm P8R2\_SOURCE\_V6_v06}$ file, respectively. The XML file uses the spectral model from the 3FGL catalog for all point sources in the field \cite{Fermi:2015}.

\begin{figure*}
\begin{center}
\includegraphics[scale=0.5]{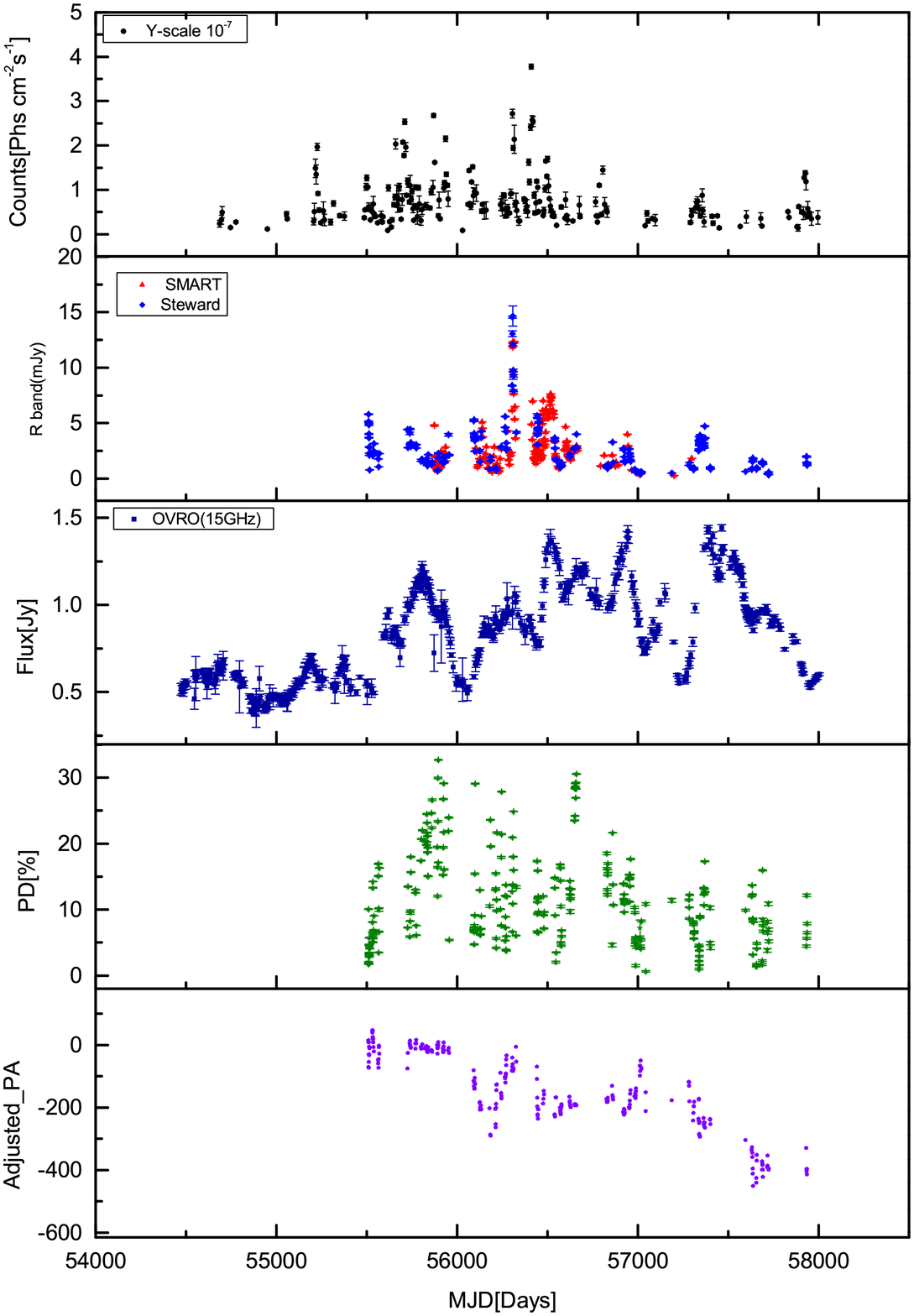}
\caption{The light curves of $0.3\to 300$ GeV $\gamma$-ray, optical R band, radio 15 GHz, polarization degree and adjusted polarization angle are plotted from up to bottom panels. }
\label{Fig:7LC}
\end{center}
\end{figure*}

 The {\it unbinned likelyhood method} is used to analyze data. We adopt the time bin to be $5$ days, which will produce a light curve with 664 data points. We considered the photon energy in the range $0.3 \sim 300$ GeV, and did not divide the energy range to obtain the energy dependent fluxes. In the 3FGL catalog, PMN J2345-1555 is named as '3FGL J2345.2-1554', which has a time averaged log-parabola spectrum $dN/dE \propto (E/E_b)^{- \alpha + \beta \log (E/E_b)}$. Parameters of sources inside the 5 degree radius from the center of ROI are left free, others lying between 5 to 15 degree radius are fixed. Note that there is another FSRQ source  PKS 2345-16 (3FGL J2348.0-1630) near the center. Distance between  PKS 2345-16 and PMN J2345-1555 is less than 1 degree. The public weekly bin aperture photometry light curves of fermi show the same light curve for these two targets. In model files, the time averaged photon flux of PMN J2345-1555 is one order larger than that of  PKS 2345-16. The discrepancy in flux diminished the contamination problem especially in light curve correlation analysis. The target information are extracted from the output files of {\it gtlike} with TS threshold $25$. The $0.1 -300$GeV flux light curve is presented in Figure \ref{Fig:7LC}.

To perform the spectra analysis, the $\gamma$-ray data is also analyzed in energy binning method. We divide the energy range $0.1 \to 243.1$ GeV into seven logrithmically equal sections. For each section, we use the $gtlike$ to obtain the fitting results. Then, parameters of the target is extracted for TS larger than 10. At least four resulted fluxes in one time bin are required to perform the linear fitting by using $\log F_{\gamma}(E)= \alpha \log E+A_0$. The procedure to some extent  can diminish the coupling effects between flux and indices in likelyhood analysis.   The linear fittings are done on those time points with at least three valid fluxes. Finally, we obtain a light curve of the $\alpha$ with 43 points and read out its corresponding flux.

{\bf Photometry and polarization data} The optical and the  polarization data are downloaded from the public Steward observatory database \footnote{http://james.as.arizona.edu/~psmith/Fermi/}, which is performed to support the {\it Fermi} missions \cite{Smith:2009}. The optical data were obtained by the $2.3$ m Bok Telescope and the $1.54$ m Kuiper Telescope. The polarization measurement was performed by the SPOL spectropolarimeter \cite{Schmidt:1992}.
  Due to various reasons like the lunar phase, the bad weather, and competition from other programs for telescope time, the data was unevenly sampled. The polarization degree (PD) and polarization angle (PA) on the web have been calibrated. In this work, both $\delta V$ and $\delta R$ relative magnitude of the target source from 2011 November 7 to 2017 July 1 are available.
The optical $R$ and  near-infrared $J$ band data from SMARTS were also downloaded \footnote{http://www.astro.yale.edu/smarts/glast/home.php}. The data is observed from 2011 Nov. 08 to 2015 Jun. 26, which has already been calibrated \cite{Bonning:2012}. Although it covers a shorter time duration than the Steward data, the more intensive sample rate makes it suitable for correlation analysis. The optical finding chart of the target presents the $R$ band magnitude
of the comparing star '6' to be 16.04, this comparing star is named as 'A' in Steward's finding chart. The foreground galactic extinction are 0.067, 0.053, 0.017 mag in $V$, $R$ and $J$ band, respectively \cite{Schlafly:1982}. We use this comparing star to calibrate the $R$ magnitudes of the Steward data.

For the PA, there is the $n\pi$-ambiguity problem, so the continuous rotation of the PA can not be directly presented by the available data \cite{Marscher:2008,Kiehlmann:2016}. We diminish the $n\pi$-ambiguity
by using the method proposed by \cite{Kiehlmann:2016}. The code is written according to the formula
\begin{equation}\label{eq:npi}
\chi_{i,\rm adj}=\chi_i -n \cdot 180^{\circ}, \qquad n={\rm round}\left( \frac{\chi_i-\chi_{i-1}}{180^{\circ}}\right).
\end{equation}
Since the original PA data covers almost six years, several observation gaps exist for the PA time series due to the observation cycles. The change speed of PA is rapid for this target, and a hand-operated $\pm180^{\circ}$ shift for one observation cycle is allowed. However, we did not adjust the data by hand to keep the smoothness of the adjustment.  The adjusted PA value are in the range $-600^{\circ} \sim 180^{\circ}$. The adjusted PA light curve is presented in Figure (\ref{Fig:7LC}). The polarization angle is derived from the $Q$ and $U$ Stokes parameters, which present more information than $\chi$. $Q$ and $U$ Data from Steward is also analyzed in the following sections.

{\bf Radio $15$ GHz data}
This research has made use of data from the OVRO 40-m monitoring program \cite{Richard:2011}. The measurements are not evenly sampled and are performed about twice per week. In this study, the observed data are collected from 2010 Nov. 9 to 2017 Sep. 8\footnote{http://www.astro.caltech.edu/ovroblazars/}, during which 570 measurements are available.  The data has already been calibrated, and light curve is presented in Figure (\ref{Fig:7LC}).

\section{Correlation analysis}

In order to investigate the location of the emission regions in jets and the emission mechanism, we perform the correlation analysis between different light curves and study the time-resolved spectra in both the $\gamma$-ray and optical bands.

\subsection{Analysis methods}

The correlations for light curves at different bands are important to reveal the emission mechanism of blazars. For unevenly sampled light curves, the discrete correlation function (DCF) is suitable to calculate correlations  \cite{Edelson1988}. However, DCF can also predict spurious peaked lags due to the intrinsic properties of the light curves \cite{Max:2014a}. Besides, the value of DCF can exceed the $[-1,1]$ range due to the few valid points at certain lags. Welsh invented the localized cross-correlation function (LCCF) to constrain the range of correlations \cite{Welsh:1999}. It is shown that LCCF is also efficient than the DCF in picking up the physical signal. Thus, we take use the LCCF to calculate correlations between light curves at different bands. In order to estimate  the significance of the time lag, the Monte Carlo (MC) method is applied to estimate the significance of the correlation. The MC procedure follows \citet{Max:2014a}. First, $10^4$ artificial light curves are simulated by the method given in \citet{Timmer:1995} (TK95).  The spectral slope of power density spectrum (PDS) is fixed at $\beta=2$.
Then LCCF between these artificial light curves and the observed light curve are calculated. At each lag bin, one has $10^4$ correlation coefficients, and $0.01$\% precision can be achieved for the significance estimation. It has been verified that the distribution of these values is of the normal type. Then the $1\sigma$, $2\sigma$ and $3\sigma$ significant levels corresponding to the $68.26\%$, $95.45\%$ and $99.73\%$ chance probabilities are obtained.

\begin{figure}
\plottwo{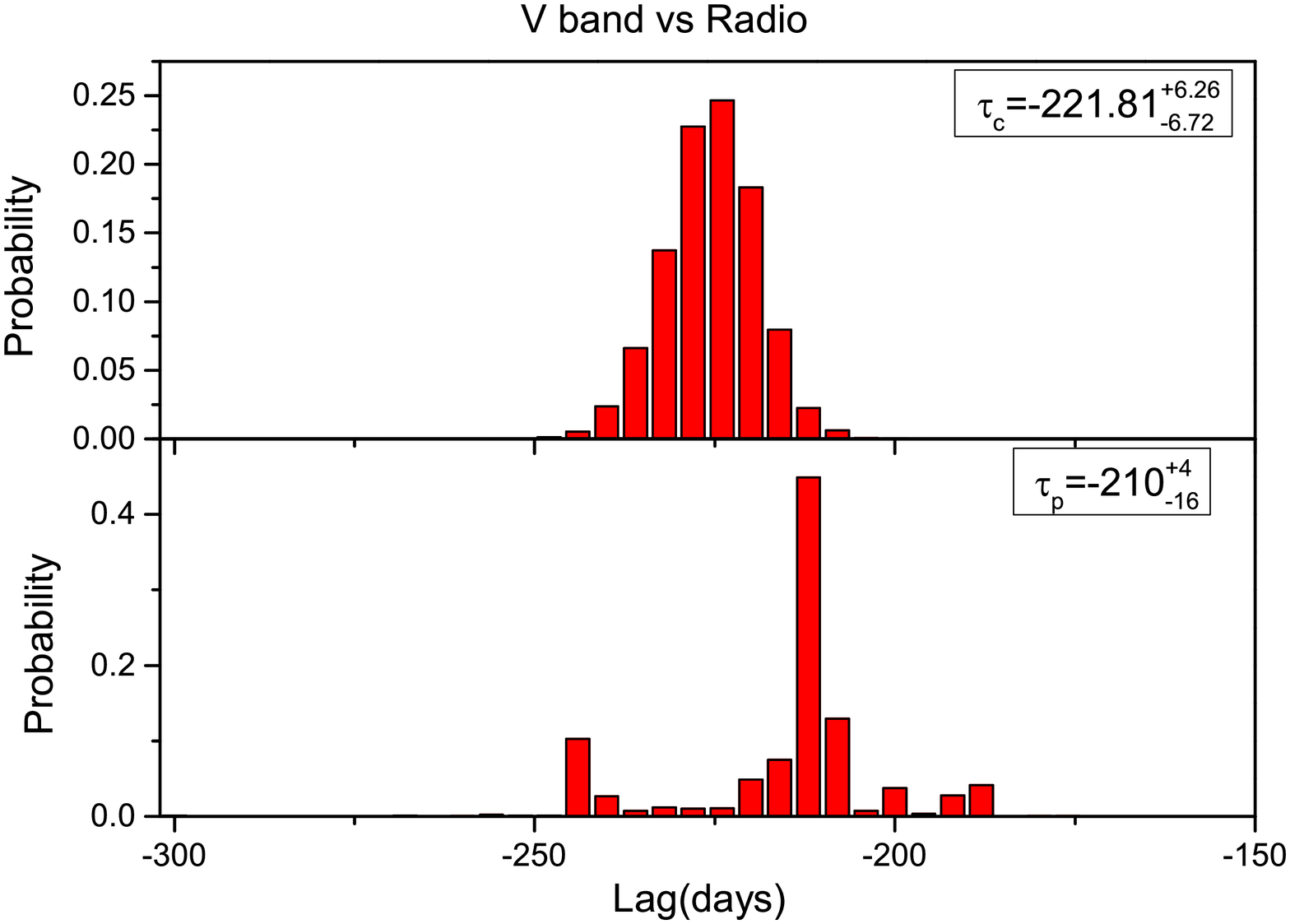}{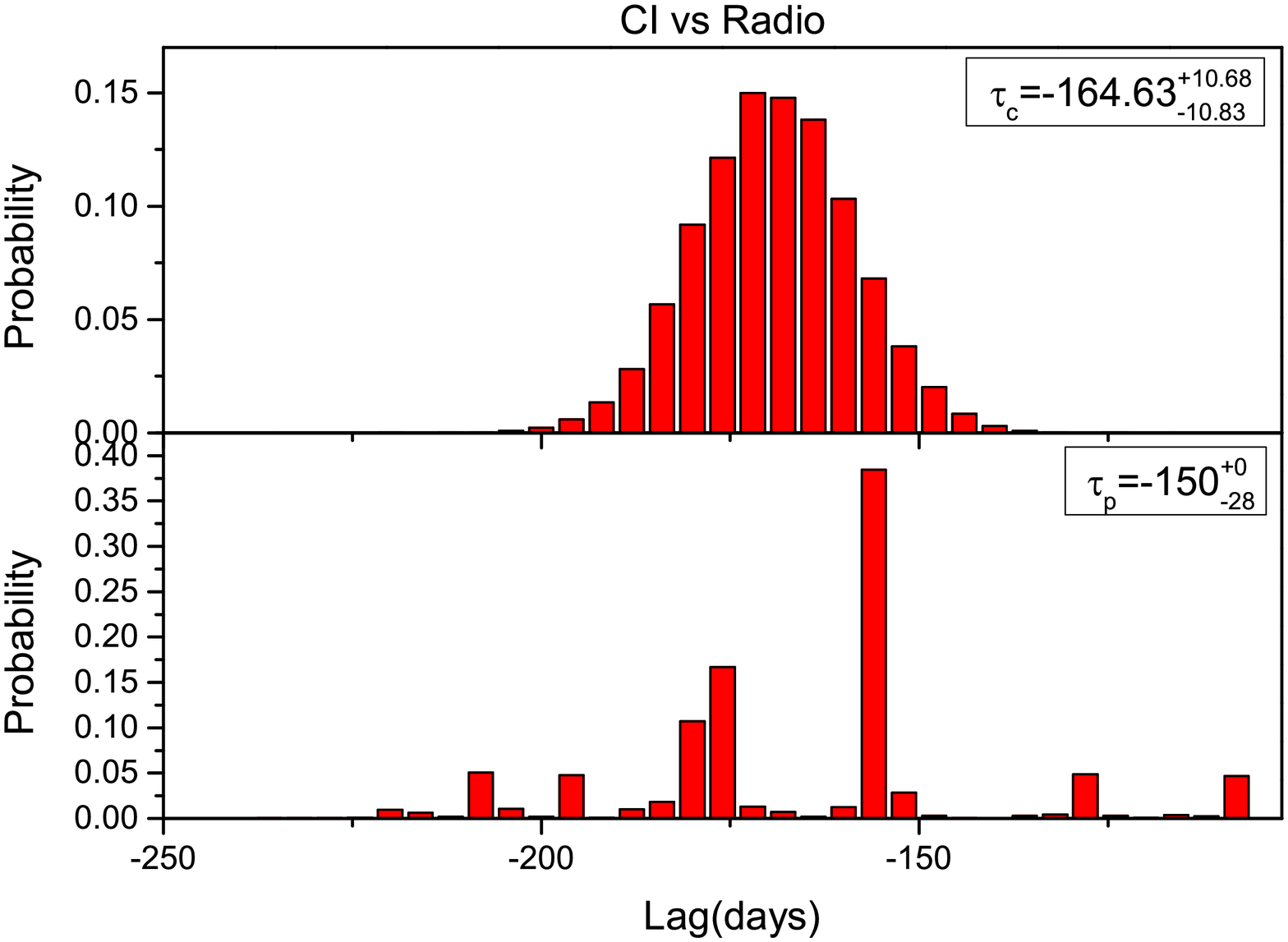}
\caption{The histogram of the centroid and peak lag times are plotted. The left two panels show the probability distribution  of FR/RSS MC at different lag times for the optical V band against the radio.
The right two panels show that for the color index against the radio. The upper and lower errors in $\tau_c$ and $\tau_p$ are the limits of $1\sigma$ range.    }
\label{Fig:Hist2}
\end{figure}

One needs to determine the time lag and its $1\sigma$ error range, if the significance level of the correlation peak is high. Two kinds of time lags can be obtained from the LCCF plots, i.e., the location of the peak of LCCF $\tau_p$ and the centroid around the peak of the LCCF $\tau_c$. The centroid lag are calculated by counting the lags whose correlation values are above half of the peak correlation.  The $1\sigma$ error ranges of both $\tau_p$ and $\tau_c$ are calculated via the model independent Monte Carlo method \cite{Peterson:1998}. We develop the code which combines the flux randomization  (FR) and the random subset selection (RSS) processes \cite{Lasson:2012}. $10^4$ time simulations are performed to obtain the distribution for both $\tau_p$ and $\tau_c$. In Figure (\ref{Fig:Hist2}), histograms of $\tau_c$ and $\tau_p$ for both the optical V-band and CI versus radio are plotted.

\subsection{Time lags}

Based on these codes, the LCCF analysis are performed between different light curves.  For  optical and radio analysis, all the highest peaks of LCCF  are beyond the $3\sigma$ significance level, see Figure (\ref{Fig:VRJlag}). Based on the these peaks, we found that the lag times for optical $V$, $R$ and $J$ bands relative to  $15$ GHz radio are  $-221.81^{+6.26}_{-6.72}$,  $-201.38^{+6.42}_{-6.02}$ and $-192.27^{+8.26}_{-7.37}$ days ($\tau_c$), respectively.
Assuming that optical leads to radio, the first peaks of the LCCF with negative lag time are at the $2\sigma$ significance level. The corresponding lag times for  $V$, $R$ and $J$ relative to  $15$ GHz  are $-65.80^{+8.67}_{-9.40}$, $-48.41^{+4.78}_{-5.65}$ and $-53.07^{+20.57}_{-27.32}$ days, respectively. Up to our knowledge, there is no method to distinguish peaks caused by periodicity and peaks caused by the distance between emission regions.
We also found that $\delta V-\delta R$ CI is strongly correlated with the radio light curve.
In  Figure (\ref{Fig:VJlag}),  it is evident that $V$ band leads to $J$ band $-10.01_{-3.22}^{+3.08}$ days. The possibility that the different sampling lead to such lag time can not be ruled out. The time lag between the optical and the near-infrared indicates that this source is very compact, and  even the optical emission is not optically thin. We tend to use the most significant peaks to calculate the distance between emission regions, and the less significant peaks are also calculated for reference.

The $\gamma$-ray is strongly correlated with $R$ band, but weakly correlated with radio.  The highest peak in the left panel of Figure \ref{Fig:gammaradio} indicates that $\gamma$-ray leads to the $15$ GHz  $103.51^{+9.57}_{-8.69}$ days, but the significance is only $\sim 1.5\sigma$. The second highest peak in this plot indicates the leading time is about 200 days. In the right panel of Figure \ref{Fig:gammaradio}, two $3\sigma$ peaks occur in the LCCF plot between $\gamma$-ray and $R$ band at lag time $-1.348^{+3.385}_{-13.502}$ and $102.581^{+4.815}_{-4.123}$, respectively. The correlation value of peak at zero lag time is a little larger than the second peak at about hundred days. This  indicates that $\gamma$-ray and optical emission region coincide. This is  possible, since a periodicity of hundred days can enhance the value of LCCF, and produce another two peaks at about $\pm100$ lag days in the plot.
Therefore, we tend to believe that the $\gamma$-ray emission region is the same with the optical. The possibility that $\gamma$-ray region lies between the optical and radio emission regions is not excluded completely.
 In Table $\ref{Table:Lag}$, we summarize the lag times (both $\tau_p$ and $\tau_c$) relative to the 15 GHz for other curves.

\begin{deluxetable}{cccccccc}
\tabletypesize{\scriptsize}
\tablecaption{Time lags relative for the radio light curve}
\tablewidth{0pt}
\tablehead{
\colhead{} &\colhead{} & \colhead{$\gamma$-ray} & \colhead{$V$ band} & $R$ band & \colhead{$J$ band} }
\startdata
Highest peaks&$t_c$ (days) & $-103.51^{+9.57}_{-8.69}$ & $-221.81^{+6.26}_{-6.72}$ & $-201.38^{+6.42}_{-6.02}$ &$-192.27^{+8.26}_{-7.37}$ \\
&$t_p$ (days)& $-106^{+28}_{-0}$ & $-210^{+4}_{-16}$ & $-208^{+23}_{-12}$  & $-187^{+2}_{-19}$  \\
\hline
Secondary peaks &$t_c$ (days) &$-203.42^{+10.18}_{-6.06}$ & $-65.80^{+8.67}_{-9.40}$ & $-48.41^{+4.78}_{-5.65}$ &$-53.07^{+20.57}_{-27.32}$ \\
&$t_p$ (days)& $-200^{+10}_{-12}$  & $-59^{+23}_{-33}$ & $-38^{+6}_{-14}$  & $-26^{+11}_{-17}$  \\
\enddata
\tablecomments{Here $t_c$ and $t_p$ denote the centroid and peak time lags (in unit of days), respectively. }
\label{Table:Lag}
\end{deluxetable}

\begin{figure}
\plottwo{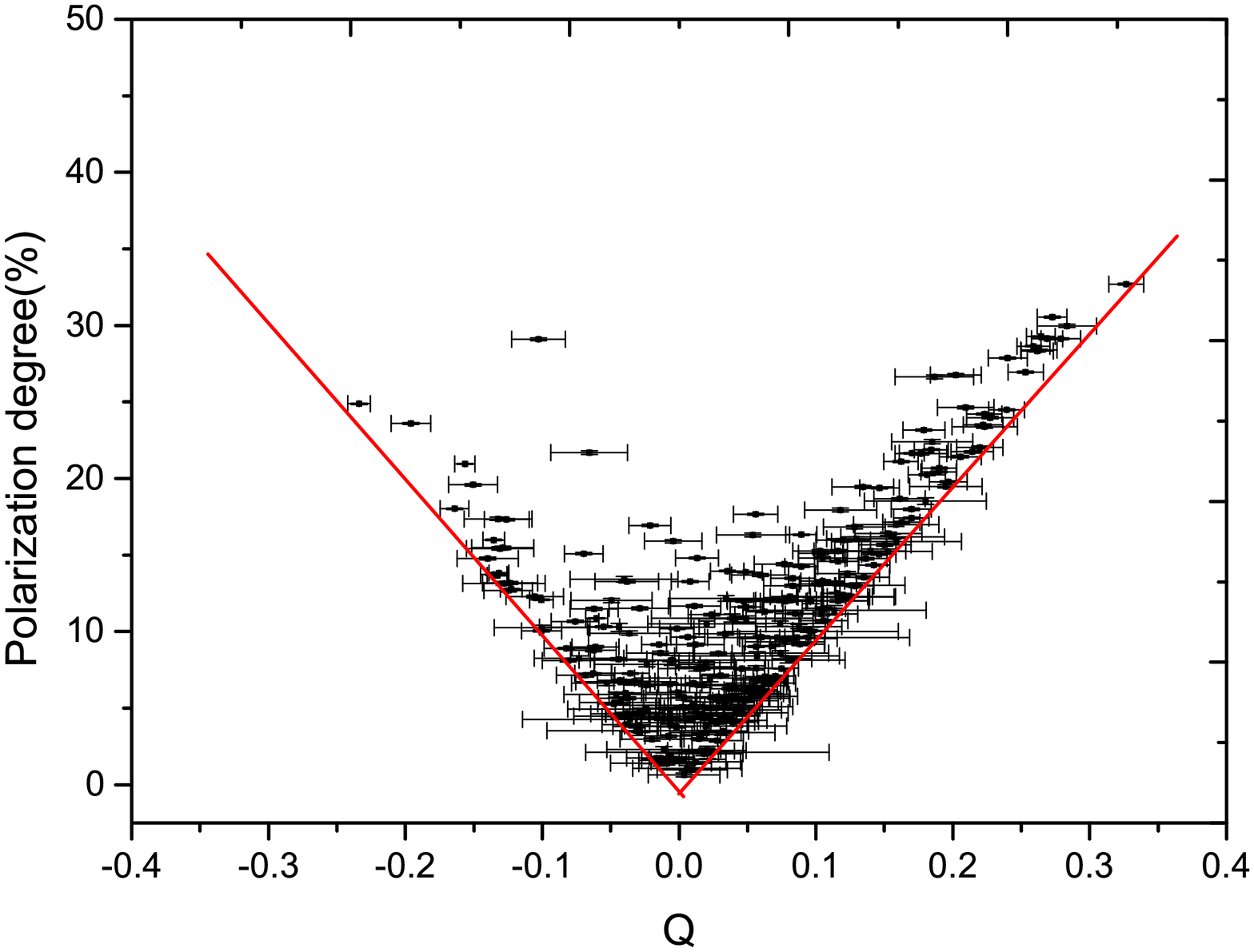}{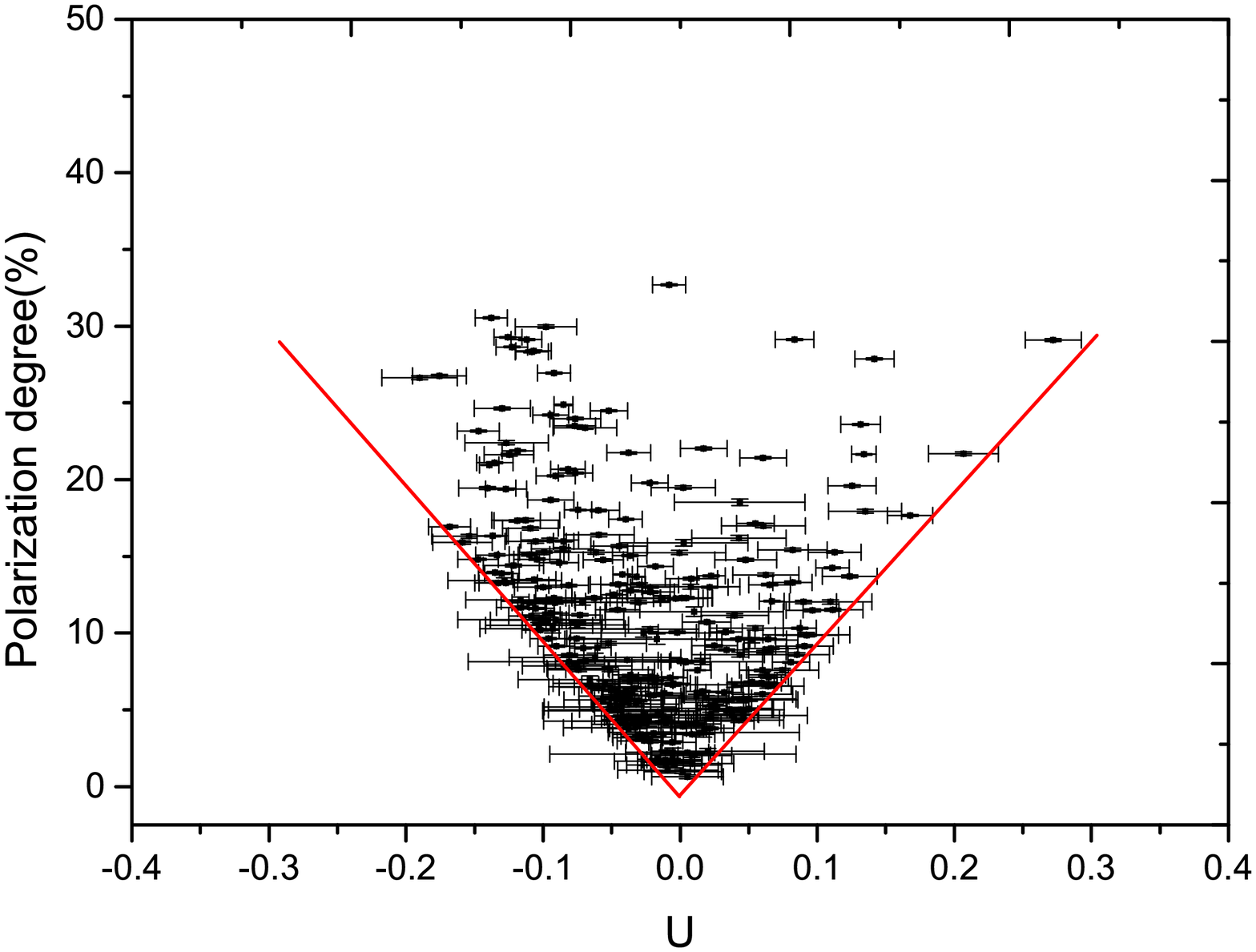}
\caption{The PD versus $Q$ and $U$ are plotted in the left and right penal, respectively. The red solid line denotes the $y=\pm 100 x$ lines.  }
\label{Fig:PDQU}
\end{figure}

\begin{figure}
\epsscale{0.8}
\gridline{\fig{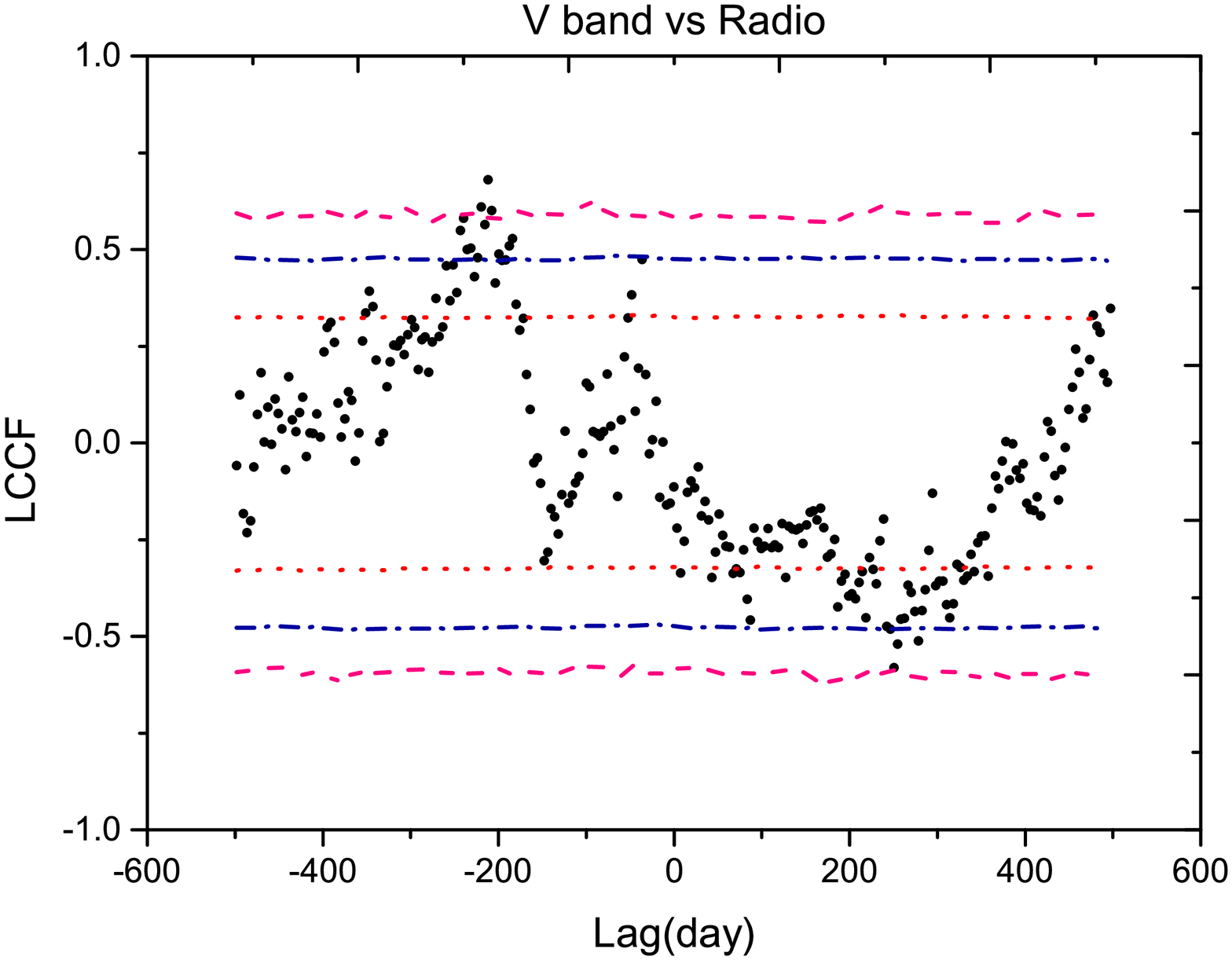}{0.4\textwidth}{(a)}
          \fig{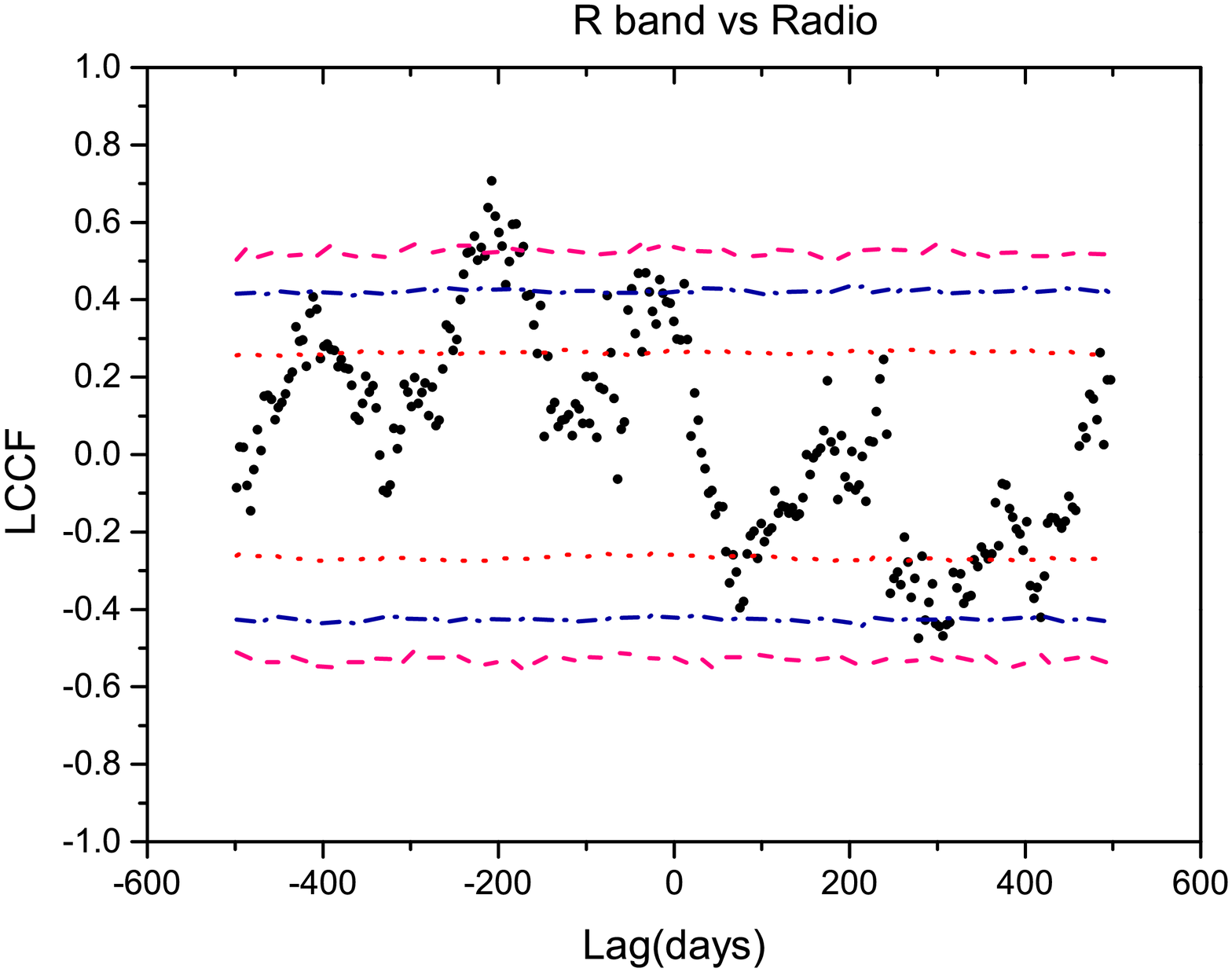}{0.4\textwidth}{(b)}
          }
\gridline{\fig{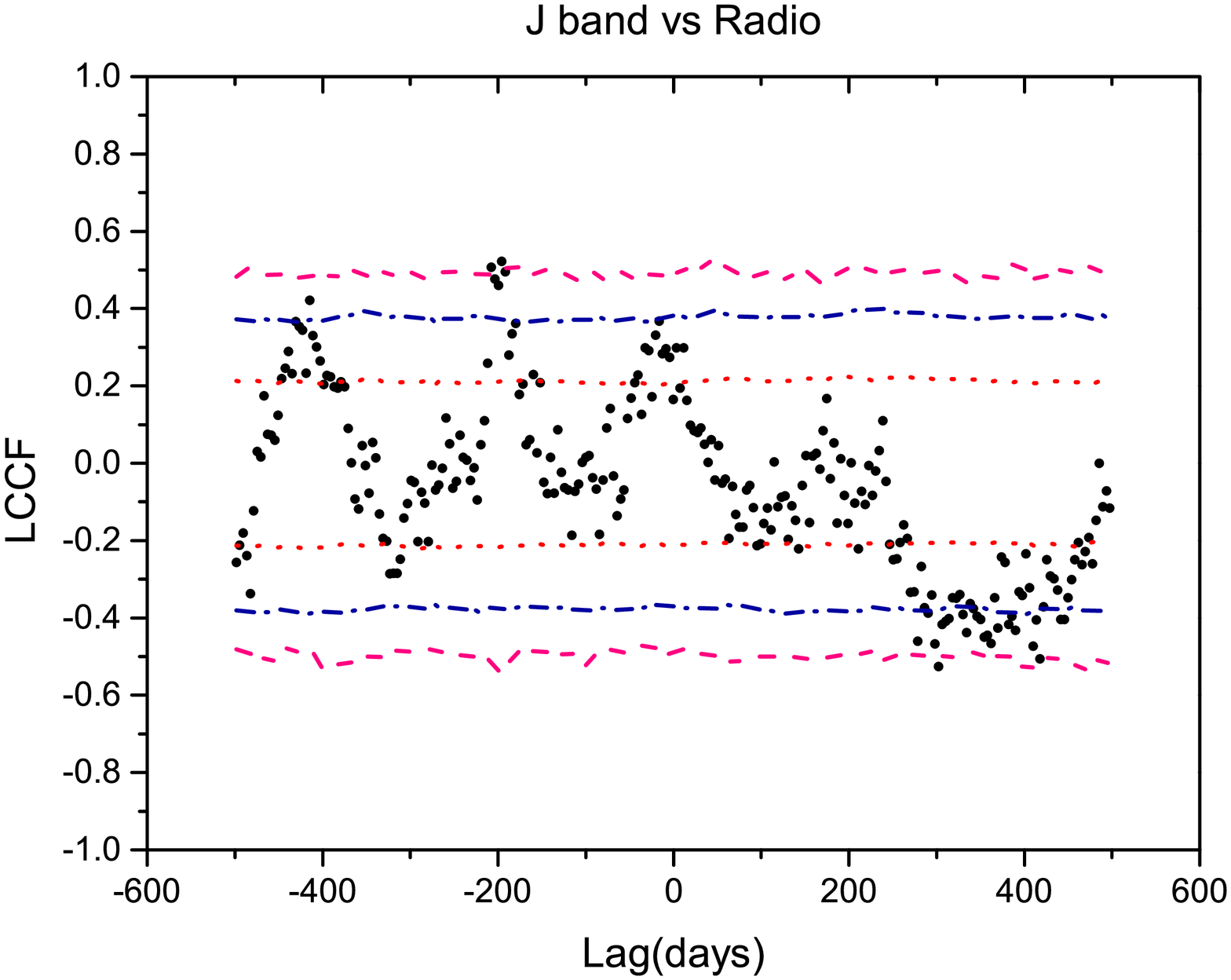}{0.4\textwidth}{(c)}
          \fig{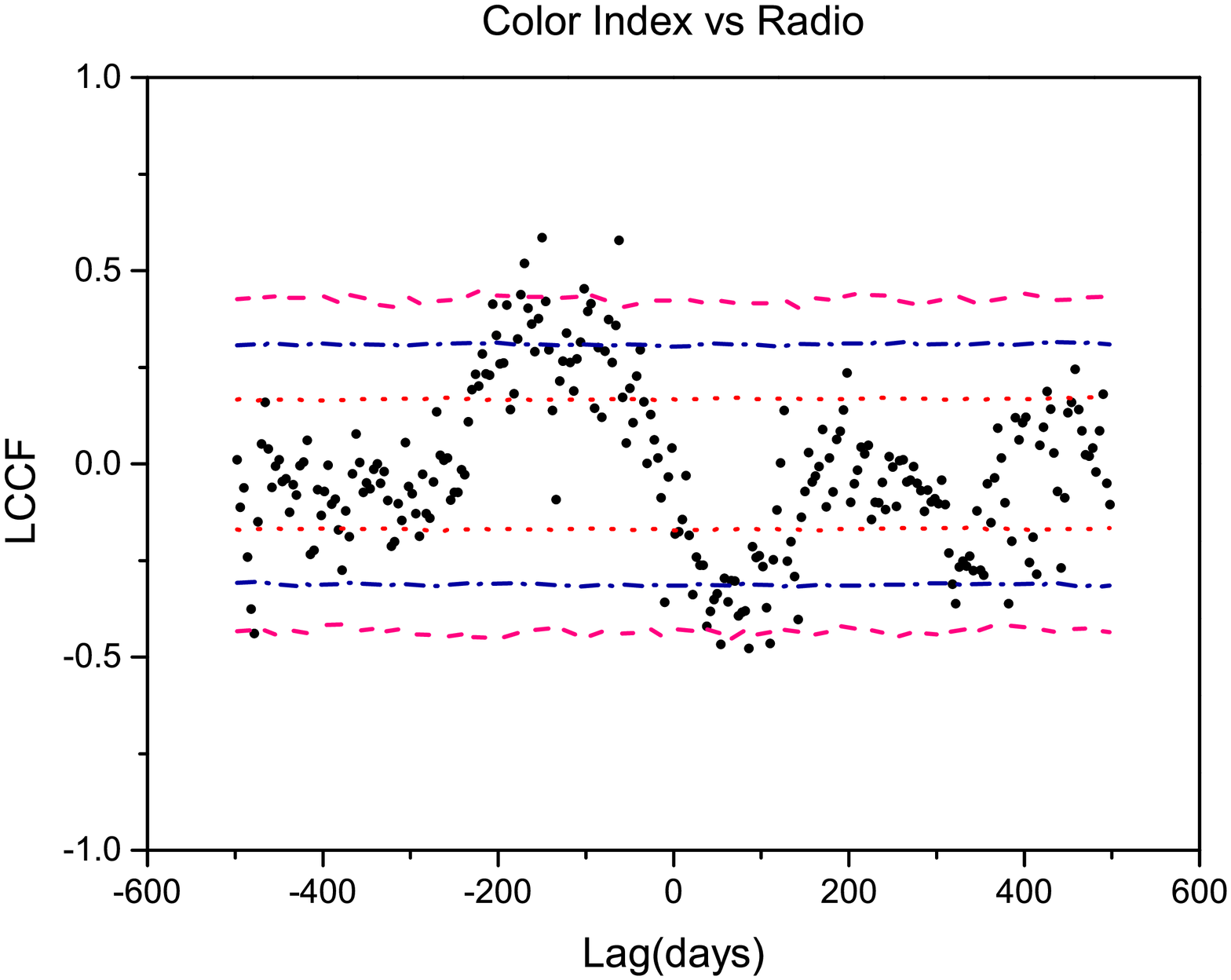}{0.336\textwidth}{(d)}
          }
\caption{The LCCFs of  optical $V$, $R$, $J$ band and $\delta V-\delta R$ CI versus 15 GHz radio light curves are plotted in a, b, c and d panels, respectively. The correlation significance are indicated by the red dot ($1\sigma$), blue dash dot ($2\sigma$) and pink  dash  ($3\sigma$) lines,  respectively. The positive value of lag time indicates the former lags behind the latter.  }
\label{Fig:VRJlag}
\end{figure}

\begin{figure}
\begin{center}
\includegraphics[scale=0.3]{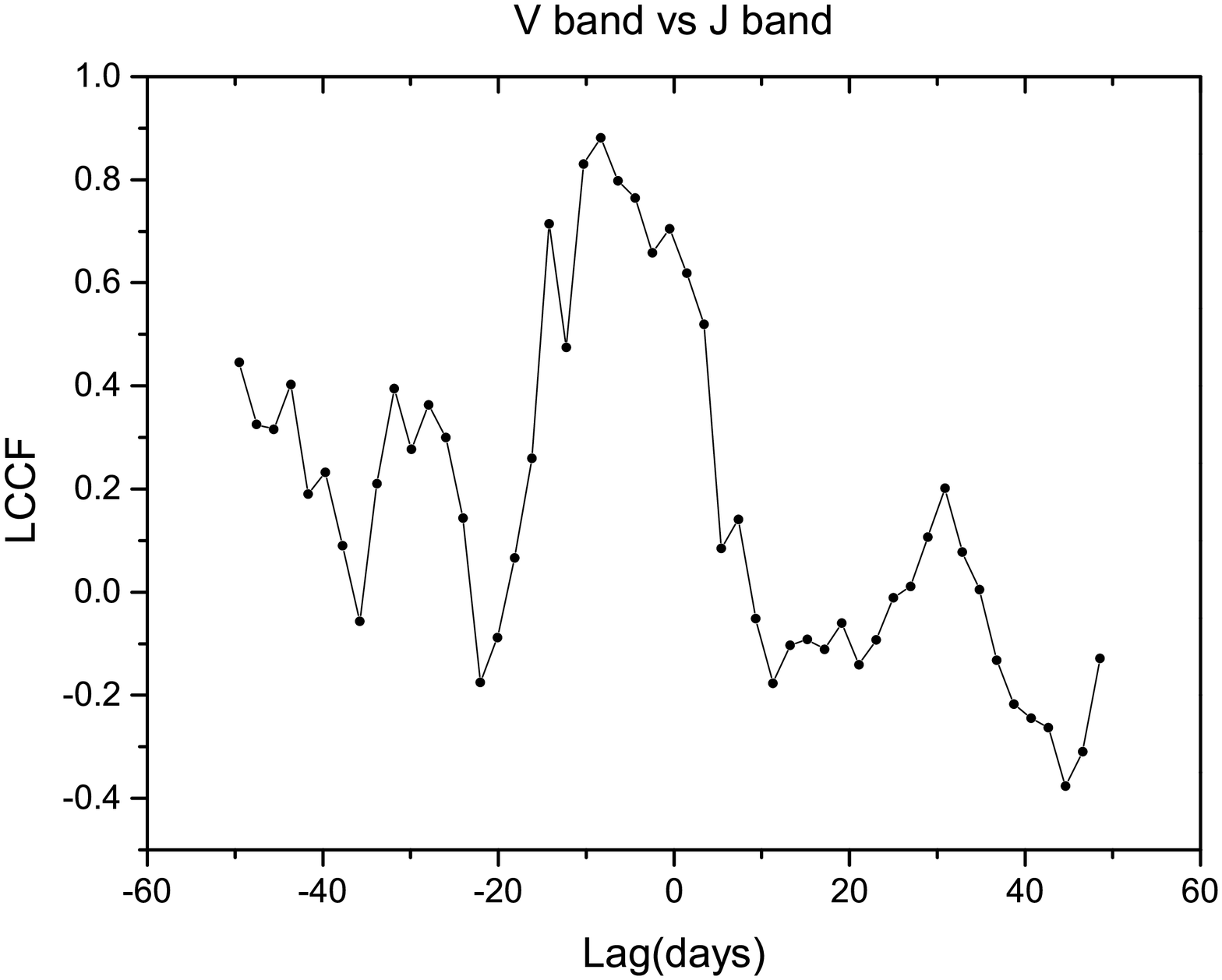}
\end{center}
\caption{The LCCF between the $V$ band and the $J$ band is plotted.} \label{Fig:VJlag}
\end{figure}

\begin{figure}
\begin{center}
\plottwo{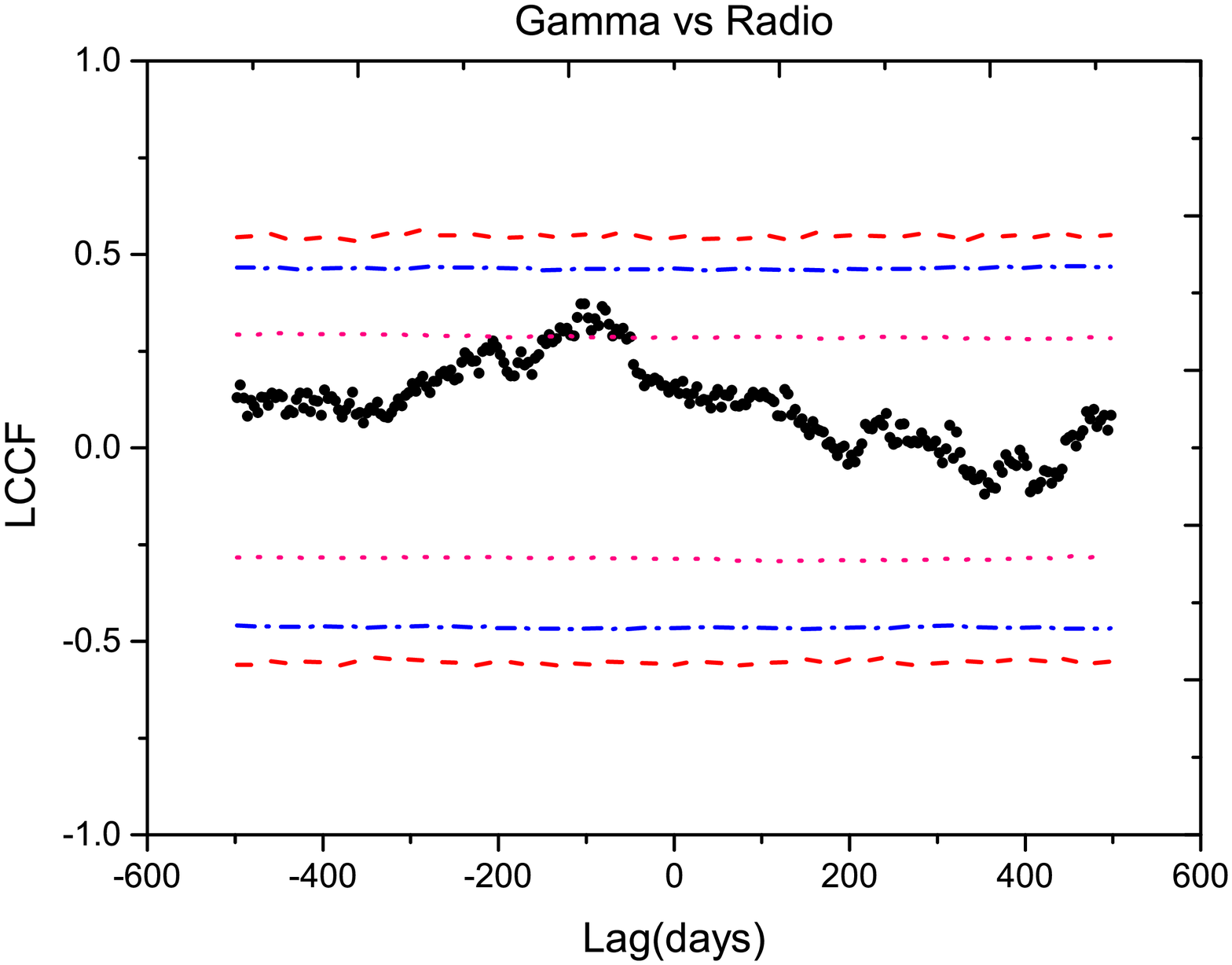}{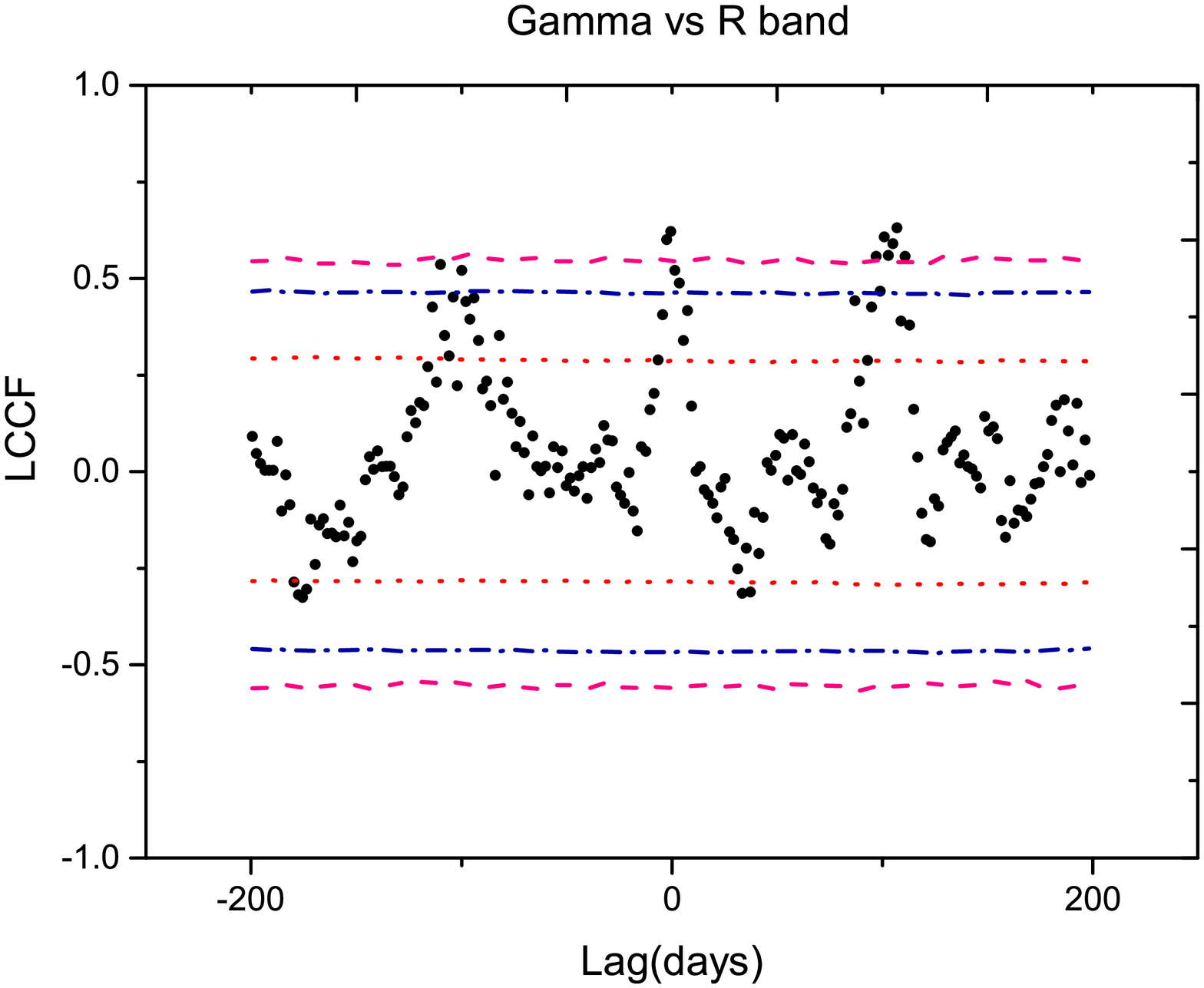}
\end{center}
\caption{The LCCF of $\gamma-$ray  versus 15 GHz and versus $R$ band   are plotted in the left and right panel, respectively.}
\label{Fig:gammaradio}
\end{figure}

\begin{figure}
\plottwo{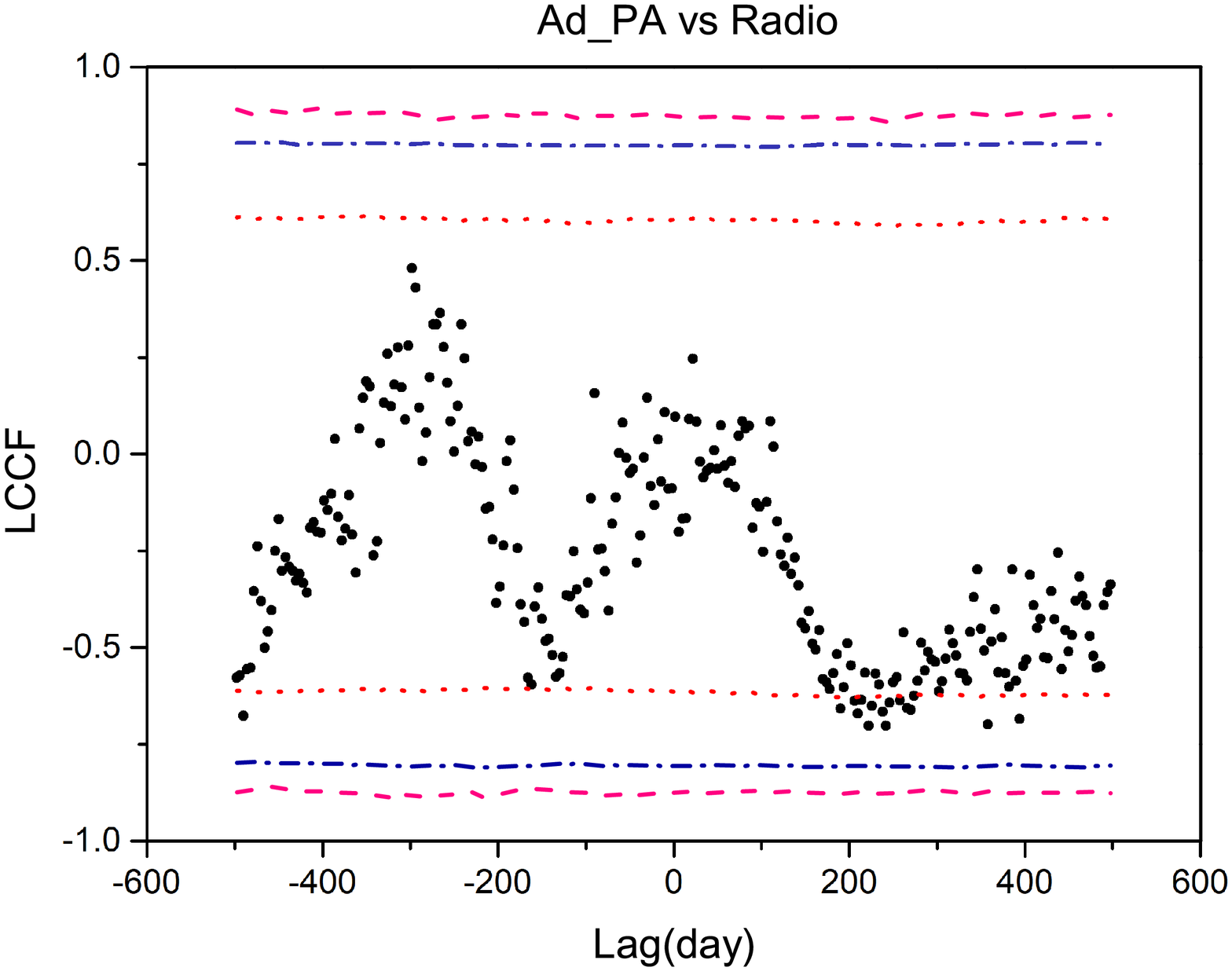}{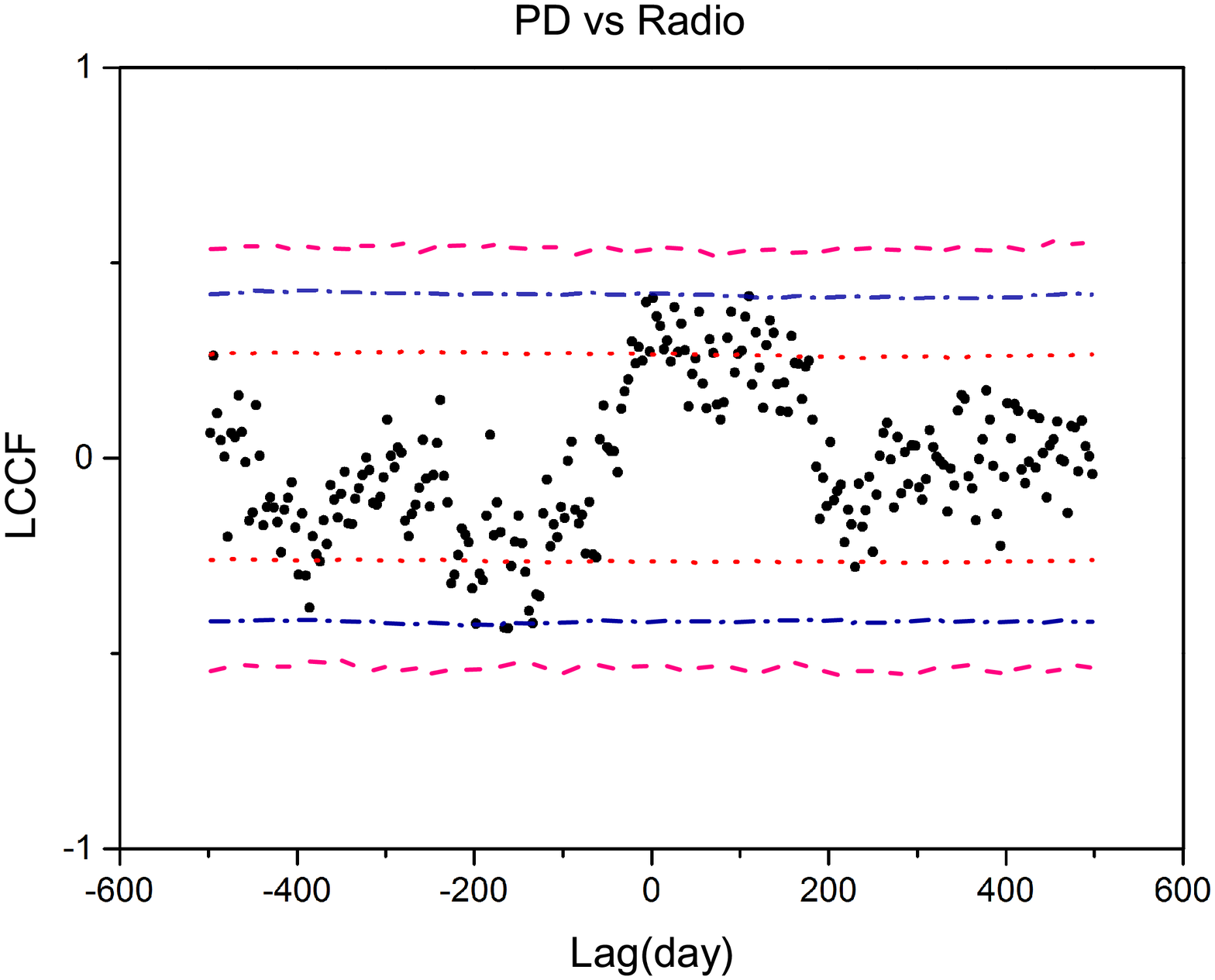}
\caption{The LCCFs of  adjusted PA and PD versus radio are plotted in the left and right panel, respectively.  }
\label{Fig:PAPD}
\end{figure}

In Figure (\ref{Fig:PAPD}), the LCCF of the adjusted polarization angle (PA) and the polarization degree (PD) versus radio are plotted. The correlation between adjusted PA and radio is  below the $1\sigma$ level. Various strategies have been tried to adjust the PA, no significant correlation is found.  The time sample of polarization is not good enough, and there is no concrete method to eliminate the $n-\pi$ ambiguity in principle. From the PD versus radio analysis, one can find $2\sigma$ correlation signals. However, the significance lines are narrow compared to the PA case. This may be caused by the moderate variation of PD light curves. At the optical lag time $-200$ days, the PD is  anti-correlated with the radio fluxes, which indicates that the PD decreases when the optical flux increases. PA is derived from the fractional Stokes parameters $q=Q/I$ and $u=U/I$ via $\chi =\frac{1}{2}\arctan (u/q)$.  $q$ and $u$ can be described by  \cite{Lytikov:2005}
\begin{equation*} \label{eq:qu}
  q=  \Pi \cos 2 \chi, \qquad u=  \Pi \sin 2 \chi,
\end{equation*}
where $ \chi$ is the PA in the observer frame. In Figure (\ref{Fig:PDQU}), it is shown that PDs are distributed inclined to the $y=\pm 100 x$ in the PD versus $q$ plot, and are distributed randomly in the PD versus $u$ plot. From equation (\ref{eq:qu}), this indicates $\cos 2\chi= \pm 1$ is preferred for high PD. This phenomenon can be explained by that the magnetic field in the emission zone is aligned in a relative fixed direction.

\subsection{Color index and spectral index}
\begin{figure}
\plottwo{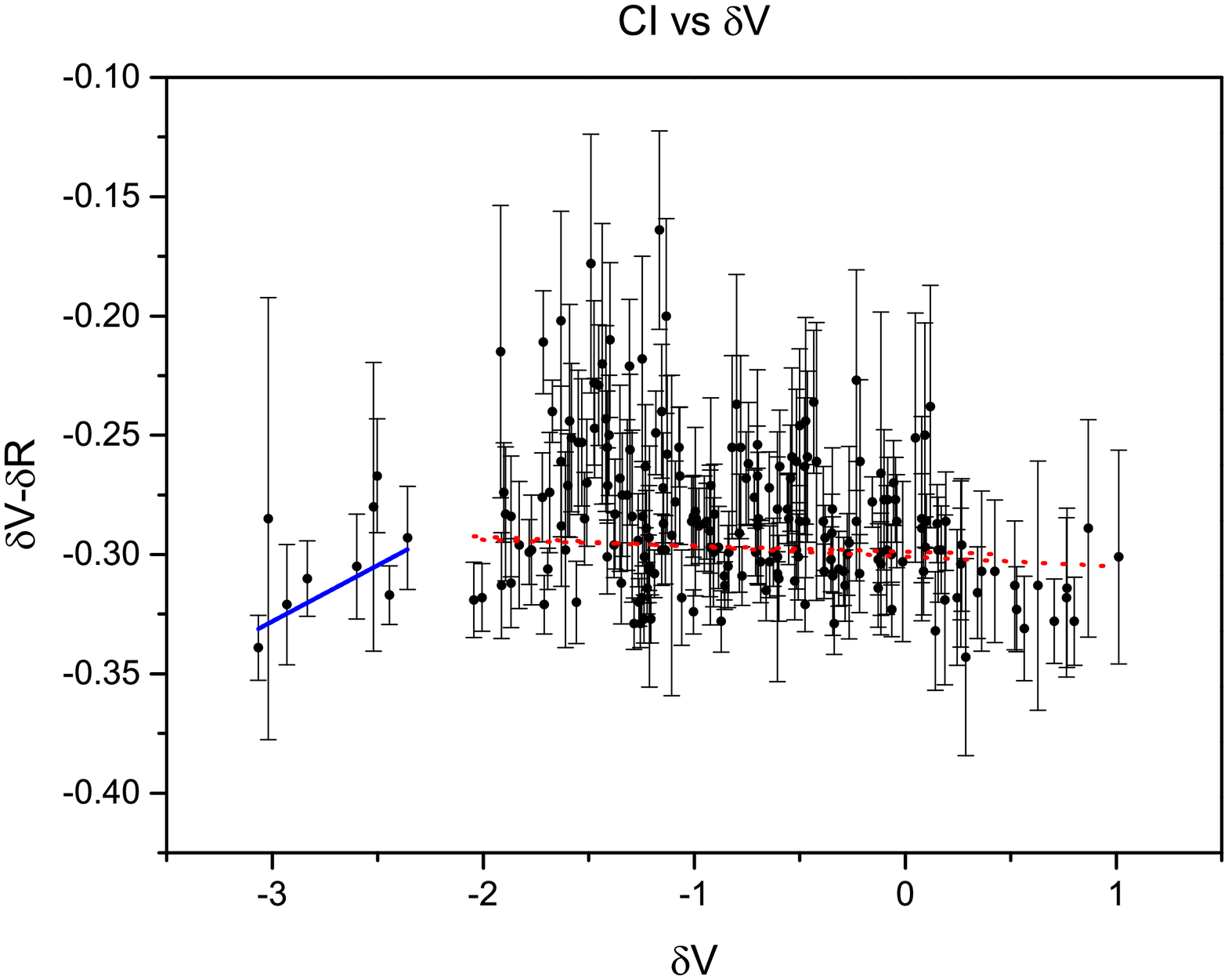}{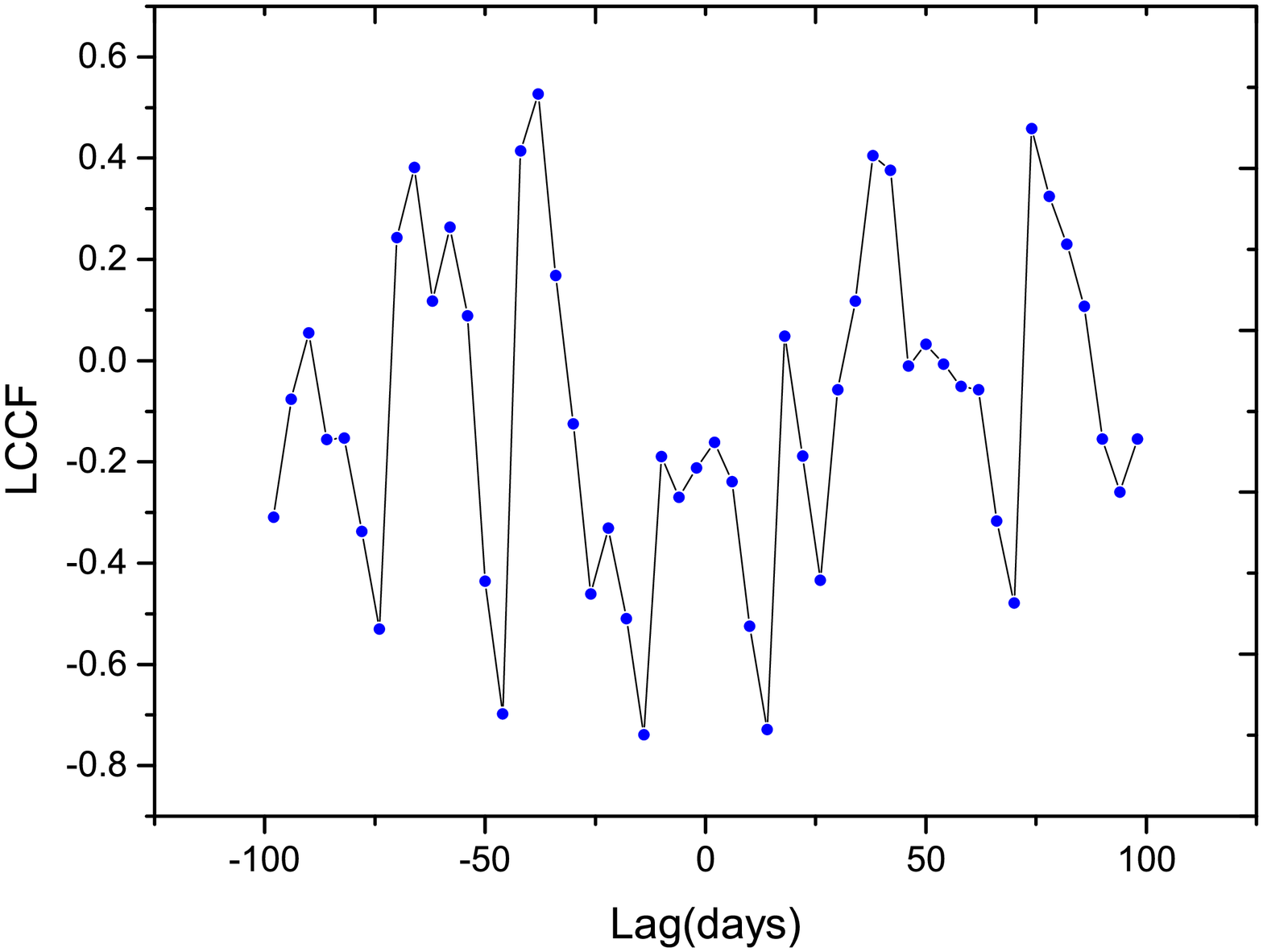}
\caption{The $\delta V-\delta R$ versus the $\delta V$  is plotted in the left panel, and the LCCF between the $\delta V-\delta R$  and $\delta V$  is plotted in the right panel. In the left panel, the linear fitting analysis $y=a+bx$ is applied to two ranges of $\delta V$. The blue solid line denotes the fitting results for the range $\delta V<-2$,
 while the red dash line denotes that of $\delta V>-2$. The slope $b$ and the Pearson's r value for the blue and red lines are (0.047, 0.642) and (-0.004,-0.107), respectively. }
\label{Fig:V_RVfit}
\end{figure}

The light curves of the spectral index and the flux can be analysed by the LCCF analysis, which offers us a time-dependent method to study the variation behavior.
To investigate the time-independent CI behaviors, we plot the $\delta V-\delta R$  versus  $\delta V$  in the left panel of Figure (\ref{Fig:V_RVfit}). It is evident that the $V$ band flux is strongly variable, since the $\delta V$ covers from $-3$ to $1$, while the $\delta V-\delta R$ CI is in the range $-0.4 $ to $-0.1$. We make use of the linear fitting $y=a+bx$ to study their relations. For $\delta V<-2$, the green (solid) line has a slope 0.047, and its Pearson's r value is 0.642. This indicates a bluer-when-brighter (BWB) trend for the flare state. For $\delta V>-2$, the ($b,r$) of the cyan (dash dot) line is ($-0.004,-0.106$). This means that the CI is almost independent of the flux variation in non-active states. In the right panel of  Figure (\ref{Fig:V_RVfit}), the negative correlation coefficients around the zero lag time indicate that $\delta V-\delta R$  and  $\delta V$ behave inversely in variation. This is a redder-when-brighter trend. The value of LCCF is in fact the Spearman coefficient, which also presents an correlation analysis of the two light curves. Around zero, the average value of LCCF is -0.2, which indicates that the redder-when-brighter (RWB) trend is not significant.

To study the variation behavior, we also extract one dimensional spectra before and after the evident flare around MJD 56306 from the Steward database. These spectra does not correct the Galactic interstellar extinction, reddening and atmosphere absorption. These corrections bring problems in calibration, but have little effects on the spectral patterns. In Figure (\ref{Fig:1dspec}), it is evident that as the flux increases, the flux  of shorter wavelength increases more evidently than the longer wavelength. This indicates a BWB trend during the flare. The slope of the continuum varies from $-0.168\pm0.006$ (linear fitting result) at MJD 56269 to $0.468\pm 0.006$ at MJD 56305. \cite{Ghisellini:2013} showed that the peak of SED locates at UVOT bands for 2013 January 14.  It was also shown that the accretion disk model fits SED well in 2008 December, and the peak of SED locates at far infrared bands at low flux state. Ghisellini et al argued that the change of the emission region from inside to outside BLR is the primary mechanism to cause the shift of the peak frequency. Since fluxes are recorded in the observer frame, the variation of Doppler factor can also shift the peak frequency. The formula of Lorentz transformations are given by \cite{Urry:1995}
\beq
\nu_p=  \delta \nu'_{p}, \qquad F_{\nu} =\delta^{3+\alpha} F'_{\nu'}.
\eeq
where $\delta=\Gamma^{-1} (1-\beta \cos \theta_{\rm obs})^{-1}$ is the Doppler factor. The variation of $\theta_{\rm obs}$  can lead to the variation of $\delta$, and hence $\nu_p$ and $F_{\nu}$. This may also affect the color index behavior. Besides, the slope of MJD 56326 is $0.589\pm 0.007$, which is higher than that of MJD 56305. The sampling of the spectra has a large gap between 56305 and 56326, we have no idea about the variation trend.

\begin{figure}
\plottwo{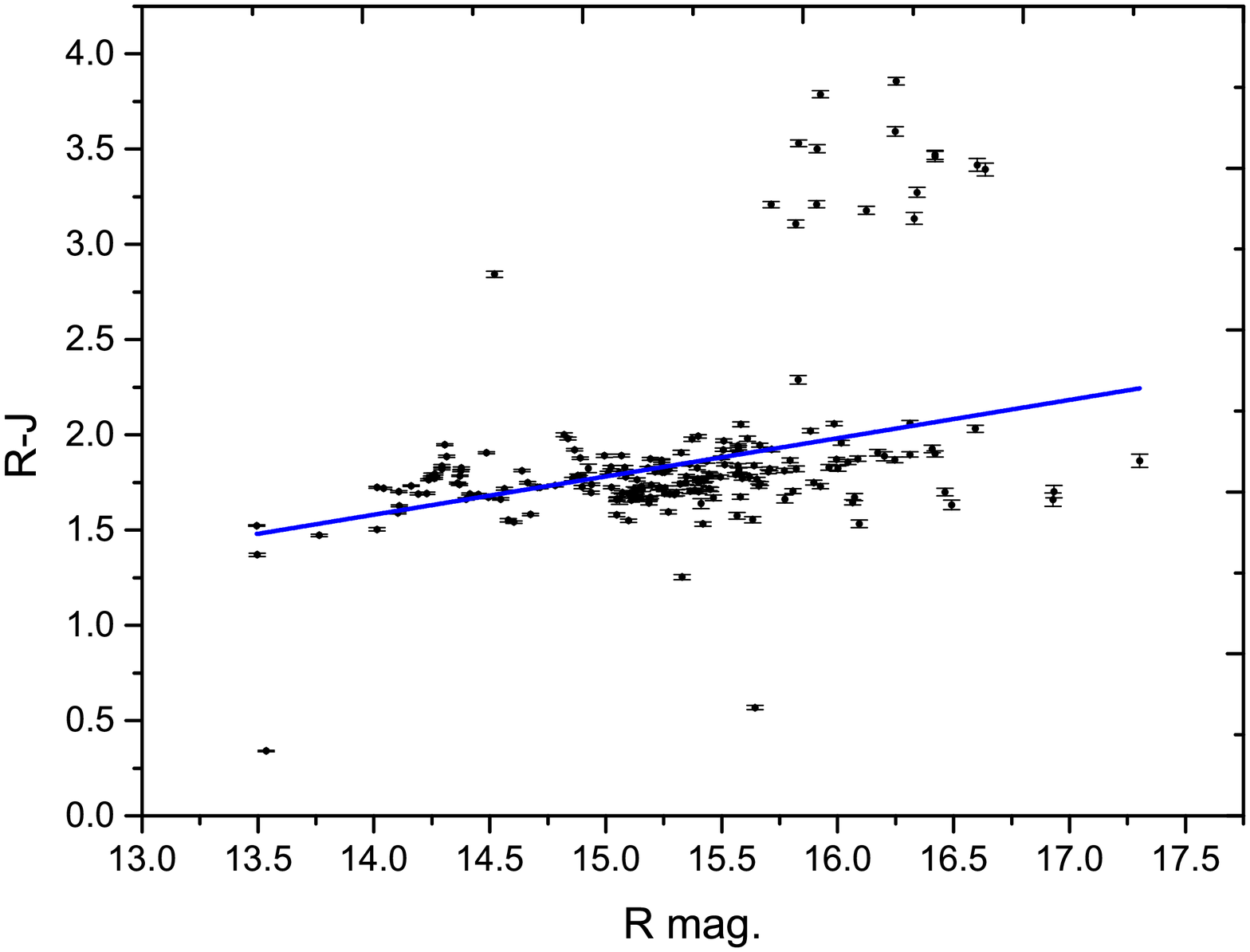}{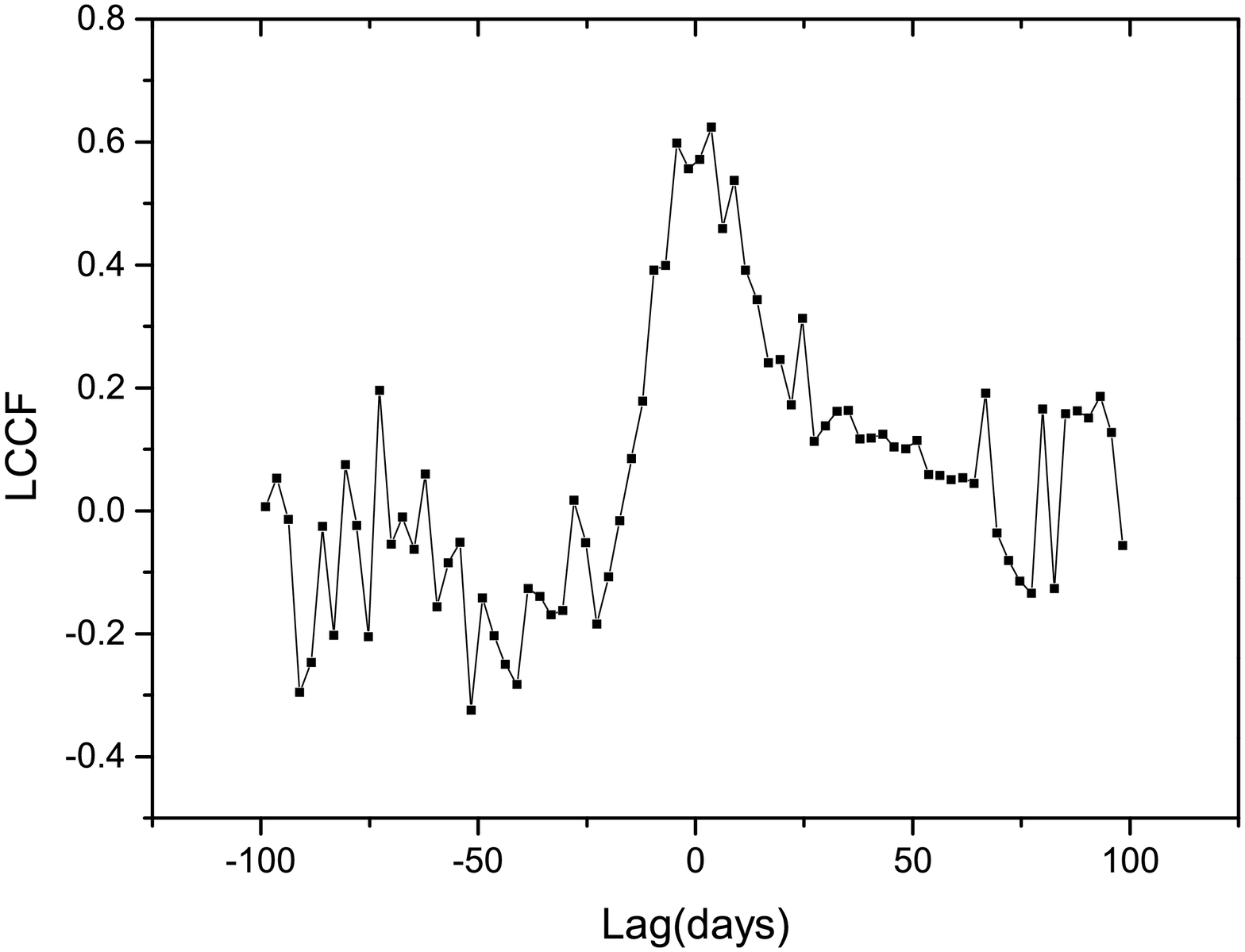}
\caption{The $R-J$ versus R magnitude and the LCCF between them are plotted in the left and right panel, respectively. The (b,r) of the blue solid line in the left panel is (0.201, 0.433).  The peak in the right panel has a correlation value 0.624. \label{Fig:RJR}}
\end{figure}

\begin{figure}
\begin{center}
\includegraphics[scale=0.3]{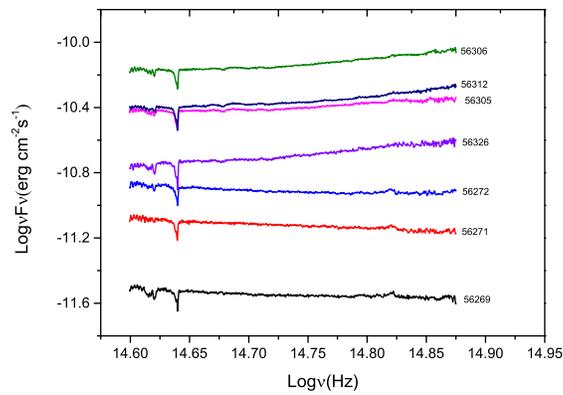}
\end{center}
\caption{The one dimensional $\nu F_{\nu}$ spectra during one flare from MJD 56269 to MJD 56326 are plotted. The flux on MJD 56306 represents the highest flux of $R$ band light curve. } \label{Fig:1dspec}
\end{figure}

In Figure (\ref{Fig:RJR}), the $R-J$ versus $R$ and LCCF between them are plotted in the left and right panel, respectively. In the left panel, the Pearson's coefficient of the linear fitting (blue solid line) is  0.433, and the slope of it is 0.201. This is a BWB trend.  In the right panel, the peak of LCCF  is 0.624, since better time sample of the data leads to a more significant correlation. Compared with the behavior of the $\delta V-\delta R$, we found that this two color index varies differently. One reason for the different behaviors of CIs may be due to the time samples. The $R$ data from  SMARTS have intensive sampling than data from Steward. The former records more flare states than the latter. The long term but sparse sample data of Steward may contain more variation components in quiescent states. The BWB trend also agrees with the spectra behavior of one extreme flare around MJD 56305.
In the quiescent state,  contributions from accretion disk and BLR region are also variable, which may cause the complex behavior of $\delta V-\delta R$.

\begin{figure}
\plotone{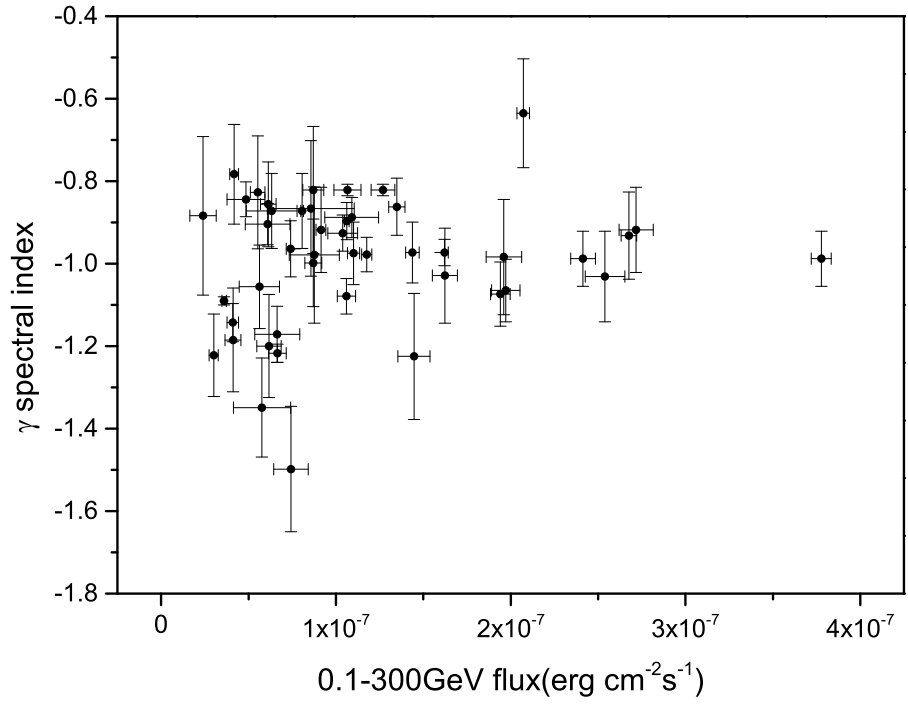}
\caption{The $\gamma$-ray spectral indices versus $0.1-300$ GeV flux is plotted.}
\label{Fig:GLag}
\end{figure}

The spectral indices of $\gamma$-ray versus  $0.1\to 300$ GeV fluxes are plotted in Figure(\ref{Fig:GLag}). We found that the spectral index tends to $-1$ at high flux states, and is much more scattering at low flux states. Such phenomena have also been reported by \citet{Lico:2014} for Mrk 421 and \citet{Algaba:2018} for 4C 38.41. It is notable that the spectra indices of three targets tend to approach certain values when the flux reaches its maximum, see  Figure 6 in \cite{Lico:2014} and Figure 5 in \cite{Algaba:2018}. One possible reason is that  the spectral index of the gamma-ray reflects that of the electron directly in the Compton scattering process. The electron spectrum is mainly determined by the acceleration mechanism in the most active state.  The three targets all are of FSRQs, but have different values for the spectral index at their maximal fluxes.
For the shock acceleration, the typical spectral slope is $p \approx 2.2\sim 2.3$ \cite{Achterberg:2001}. For the magnetic reconnection, the resulted spectrum of electrons can be very hard, and an extreme  $p \approx 1$ can be achieved \cite{Guo:2015}.


\section{Theoretical frame}
The strong correlations between the optical, near-infrared and radio indicate that the jet in PMN J2345-1555 is very compact. Due to the opacity, the photosphere positions at sequent frequencies will systematically shifts towards the downstream of the jet, this leads to the core-shift phenomenon in the milliarcsecond resolution VLBI observations \cite{Konigl:1981,Lobanov:1998,Hirotani:2005}. \citet{Kudryavtseva:2011} proposed that time lags between different frequencies can play the same role as the core shift since they are both related to the projected distance in the sky. By using the time lags, many jet properties can be studied. In the following, we will present the theoretical frame. Based on which, many variation phenomena can be explained. Then, we explore the location of the optical emission regions, and derive many aspects of the jet.

The convention in this part mostly follows \citet{Hirotani:2005} (H05 hereafter). First one takes the assumption of the scaling law that the magnetic field and the proper number density of electrons follow $B^*=B_1 r^{-m}$ and $N^*=N_1 r^{-n}$, where $r$ is in parsec units which represents the distance in jet, $B_1$ and $N_1$ refers to corresponding values at $1$ pc. $r_1=1{\rm pc}$ is in convention.
The flux intensity for the uniform slab of plasma in the jet is given by \cite{Hirotani:2005}
\beq  \label{eq:fluxint}
S_{\nu}=\frac{\pi}{4}\left(\frac{\theta_d}{\rm rad}\right)^2 \left(\frac{\delta}{1+z}\right)^{1/2} A \nu^{5/2}(1-e^{-\langle\tau_{\nu}\rangle}),
\eeq
where $A=(3/2)^{-\alpha}\frac{e a(\alpha)}{c C(\alpha)}(\frac{e}{2\pi m_e c})^{-3/2}B^{-1/2}$, and $\langle\tau_{\nu}\rangle$ is the geometrical averaged optical depth at $\nu$. Here $\alpha$ is a typical spectral index defined as $S_{\nu} \propto \nu^{\alpha}$. For flux in the optical thin region, we assume $\alpha<0$. The flux density here is valid both in the optically thin and thick regions, one can check that the spectral index for $\nu$ is in the range $(5/2, \alpha)$. The averaged optical depth can be expressed by $\langle\tau_{\nu}\rangle= \alpha_{\nu} R_{\rm eff}$, where $\alpha_{\nu}$ and $R_{\rm eff}$ stands for the synchrotron self-absorption (SSA) coefficient and the effective length along the line of sight in the observer frame, respectively. In geometry, one has $R_{\rm eff}=f r_{\perp} /\sin \varphi$ and $r_{\perp}=r_1 r \sin \theta$, where $\theta$ is the jet opening angle, $\varphi$ is the viewing angle, and $f$ is a dimensionless geometrical factor from integration. For the slab geometry, one can take $f=1$ and for others one usually has $f<1$. $\alpha_{\nu}/\sin \varphi$ is Lorentz invariant, The absorption coefficient $\alpha^*_{\nu}$ is given by
\beq
\alpha^*_{\nu}=C(\alpha)\frac{e^2}{m_e c}\frac{-2\alpha}{\gamma^{2\alpha}_{\rm min}} \left(\frac{e}{2\pi m_e c}\right)^{\epsilon}N^* B^{\epsilon} \nu^{*(-1-\epsilon)}.
\eeq
where $\epsilon \equiv 3/2-\alpha$ and $C(\alpha)$ originates from SSA counting (details are referred to Table 1 in H05). Considering the scaling law, the averaged optical depth  is given by
\beq
\langle\tau_{\nu}\rangle= \Sigma(\alpha) N_1 B^{\epsilon}_1 \left(\frac{\delta}{1+z}\right)^{\epsilon} r^{1-n-m\epsilon} \nu^{-1-\epsilon},
\eeq
where $\Sigma(\alpha)\equiv f r_1 C(\alpha)\frac{-2\alpha}{\gamma^{2\alpha}_{\rm min}}\frac{e^2}{m_e c} (\frac{e}{2\pi m_e c})^{\epsilon} \frac{\sin \theta}{\sin \varphi} $.    Here $\delta$ is the Doppler factor and $z$ is the redshift.
Setting $\langle\tau_{\nu}\rangle=1$, one obtains the distance between the core position for observed frequency $\nu$ and the base of the jet, i.e. \cite{Pushkarev:2012,Fuhrmann:2014}
\beq \label{eq:rcore}
r_{\rm core}(\nu)= \Sigma(\alpha)^{\frac{1-k_b}{k_r}} \left(\frac{\delta}{1+z}\right)^{\frac{k_b}{k_r}} N_1^{\frac{1-k_b}{k_r}} B_1^{\frac{k_b}{k_r}} \nu^{-\frac{1}{k_r}},
\eeq
where indices are defined by
\begin{align}\label{eq:para33}
  k_b&\equiv \frac{3-2\alpha}{5-2\alpha}, \\
  k_r &\equiv  \frac{(3-2\alpha)m+2n-2}{5-2\alpha}.
\end{align}
$r_{\rm core}(\nu)$ denotes the photosphere position of frequency $\nu$ in the jet. Assuming there are two different frequencies
$\nu_1$ and $\nu_2$, one can obtain the photosphere distance between them by two kinds of observations. One is the high spatial resolved VLBI observation, another is the time lag analysis between well sampled time series. Here we present the relation between them via the projected distance $\Delta r_{\rm proj}$ (in units of pc), i.e.,
\begin{align} \label{eq:core}
\Delta r_{\rm proj}=&[r_{\rm core}(\nu_2)-r_{\rm core}(\nu_1)] \sin \varphi \nonumber \\
              = & \frac{\Omega_{r\nu}}{r_1}\left( \nu_2^{-1/k_r}-\nu_1^{-1/k_r}\right) \nonumber \\
              =& \frac{\beta_{\rm app} c \Delta t^{\rm obs}_{\nu_1 \nu_2}}{r_1(1+z)},
\end{align}
where $\Omega_{r\nu}$, $\beta_{\rm app}$ and $\Delta t^{\rm obs}_{\nu_1 \nu_2}$ are the core-position offset, the apparent velocity of the jet and the time lag between light curves of $\nu_1$ and $\nu_2$ in observer frame, respectively. Note that $\beta_{\rm app}\equiv \beta\sin\varphi(1-\beta \cos \varphi)^{-1}$ can also be expressed as $\beta_{\rm app}=\Gamma \delta \beta \sin \varphi$, where $\Gamma$ is the Lorentz factor and $\delta=\Gamma^{-1}(1-\beta \cos \varphi)^{-1}$ is the Doppler factor \cite{Pushkarev:2010,Max:2014a,Max:2014b}. These parameters can be measured by the VLBI observation  \cite{Hovatta:2009,Pushkarev:2010,Pushkarev:2012}.
The core-position offset $\Omega_{r\nu}$ is defined as \cite{Lobanov:1998}
\beq
\Omega_{r\nu}\equiv 4.85\times10^{-9}\frac{\Delta r_{\nu_1 \nu_2} D_L}{(1+z)^2}\frac{\nu_1^{1/k_r}\nu_2^{1/k_r}}{\nu_2^{1/k_r}-\nu_1^{1/k_r}} {\rm pc}\cdot{\rm GHz},
\eeq
where $\Delta r_{\nu_1 \nu_2}$ is units of milliarcseconds, and can be measured by the core shift in VLBI radio images.
In Equation(\ref{eq:core}), it is evident that $\Omega_{r\nu}$ corresponds to the lag time $\Delta t^{\rm obs}_{\nu_1 \nu_2}$ in the following
\beq \label{eq:omegat}
\Omega_{r\nu}= \frac{ \beta_{\rm app} c \Delta t^{\rm obs}_{\nu_1 \nu_2}}{1+z}\frac{\nu_1^{1/k_r}\nu_2^{1/k_r}}{\nu_2^{1/k_r}-\nu_1^{1/k_r}},
\eeq
which enable us to measure the core-position offset via the lag time $\Delta t^{\rm obs}_{\nu_1 \nu_2}$.

From Equation (\ref{eq:rcore}) and (\ref{eq:core}), it is possible to estimate $B_1$ and $N_1$ via multiple frequencies core-shift or time lag observations in principle. The index $k_r$ can be obtained  by the core-shift measurement $\Delta r_{\nu_1 \nu_2}$ \cite{OSullivan:2009}, or by time lags $\Delta t_{\nu_1 \nu_2}^{\rm obs}$ in multiple radio band observations \cite{Fuhrmann:2014}.  Under the assumption that the magnetic filed energy density is equal to the particle, and jet is optically thick due to SSA, $k_r$ is roughly equal to 1. In the equipartition condition, the spectral index $\alpha$ is $-0.5$. Then, the magnetic field and electron number density  at 1 pc can be calculated by \cite{Hirotani:2005,OSullivan:2009}
\begin{align}\label{eq:b1n1}
  B_1 & \simeq 0.025 \left[ \frac{\sigma_{\rm rel} \Omega_{r\nu}^3 (1+z)^2}{\theta \sin^2 \varphi  \delta^2}\right ]^{1/4}, \nonumber \\
  N_1 & \simeq 3.3 \left[ \frac{  \sigma_{\rm rel} \Omega_{r\nu}^3 (1+z)^2}{\gamma_{\rm min}^2\theta \sin^2 \varphi  \delta^2}\right]^{1/2},
\end{align}
where $\gamma_{\rm min}$ is the minimal electron Lorentz factor, $ \sigma_{\rm rel}$ is the ratio between the energy density of  magnetic field and that of non-thermal particles which is set to be 1. Substituting Equation (\ref{eq:omegat}) into (\ref{eq:b1n1}), one can obtain these two parameters from the time lag result directly. The validity of these equations are based on the assumption that photosphere emission is dominant at both $\nu_1$ and $\nu_2$.

\subsection{Location of emission regions}
For this source, the jet parameters are measured by the MOJAVE projects \cite{Pushkarev:2012}.  Its redshift $z=0.621$ \cite{Shaw:2009}. The black hole mass in our target can be estimated by the measured H$\beta$ or MgII lines \cite{Shaw:2009}. The FWHMs of H$\beta$ and MgII are $3835$ and $3630$ km $s^{-1}$ in the rest frame, respectively. By the empirical relation \cite{Shen:2011,Shaw:2009}, one obtains
\beq
\log (\frac{M_{\bullet}}{M_\odot})=0.505+0.62\log \left(\frac{\lambda L_{\lambda}}{10^{44} {\rm erg \, s^{-1}}}\right)+2\log \left(\frac{\rm FWHM_{MgII}}{ {\rm km \,\, s^{-1}}}\right) \approx 8.44,
\eeq
where $\lambda L_{\lambda}=2\times10^{45}{\rm erg \, s^{-1}}$ is considered.
The apparent velocities $\beta_{\rm app}$ has been measured for three features, i.e., $7.4\pm1.2$, $4.39\pm 0.61$  and $0.91\pm 0.56$, respectively \cite{Lister:2019}. Here, we take the fastest feature with $\beta_{\rm app}=7.4\pm1.2$ for calculation. The minimal Lorentz factor of the jet is given by $\Gamma_{\rm min}=\sqrt{\beta_{\rm app}^2+1}$. Thus, we take a moderate value $\Gamma=10$ for calculation. The  radio flux variation can also predict the Doppler factor via brightness temperature $D_{\rm var}=(T_{\rm b,var}/T_{\rm b,int})^{1/3}$. while $T_{\rm b,var}=1.548\times 10^{-32}\Delta S_{\rm max}d_{L}^2\nu^{-2}\tau^{-2}(1+z)^{-1}$ and $T_{\rm b,int}=5\times 10^{10}$K \cite{Hovatta:2009}. For the target, we take $\Delta S_{\rm max}=1.5$ jansky, $\nu=15$ GHz, $\tau=100$ days and $d_{L}\approx10^{26}$ meters, and derive that $D_{\rm var}\approx 20=2\Gamma$, which is almost the maximal value for Doppler factor with $\Gamma=10$ \cite{Urry:1995}. Then, one obtains the variable Lorentz factor and viewing angle as
\beq
\Gamma_{\rm var}=\frac{\beta_{\rm app}^2+D_{\rm var}^2+1}{2D_{\rm var}}\approx 11.5, \qquad \varphi_{\rm var}=\arctan \frac{2\beta_{\rm app}}{\beta_{\rm app}^2+D_{\rm var}^2+1}\approx 2^{\circ}.
\eeq
From formula of $\beta_{\rm app}$, the viewing angle $\varphi_{\rm app}$ can be solved analytically by
\beq
 \varphi_{\rm app} =\arccos \left[\frac{\beta_{\rm app}^2-\sqrt{\beta_{\rm app}^2(\beta^2-1)+\beta^2}}{\beta(1+\beta_{\rm app}^2)}\right].
\eeq
For $\beta_{\rm app}=7.4$, the parameter in the square root must be positive, which sets a strict lower limit for $\beta$, i.e., $\beta \ge 0.99$. $\Gamma=10$ gives $\beta^2=0.99$. Approximately, the viewing angle can be estimated as $\varphi \approx 11.8^{\circ}$.  Another crude method to estimate the viewing angle is $\varphi \approx \sin^{-1}(1/\Gamma_{\rm min})\approx 7.7^ {\circ}$ \cite{OSullivan:2009}. The variation of viewing angle is an important ingredient to the variable phenomenon of blazars. The radiative blob may have a helical trajectory in jet, its moving direction oscillates. This leads to a changing of viewing angle naturally. From the MOJAVE program homepage, the moving of knot shows a possibly curved trajectory\footnote{http://www.physics.purdue.edu/astro/MOJAVE/} \cite{Lister:2019}. If the optical and $\gamma$-ray emission blob has a similar trajectory, the derived distance in Equation(\ref{eq:core}) becomes an upper limit. The pitch angle of the helical motion should be considered in the distance calculation. However, the method to measure this parameter is unknown for us currently. The helical jet will be studied in the future. Since $\varphi_{\rm var}$ mostly reflects the minimal angle between moving direction of blob and the line of sight (LS), while $ \varphi_{\rm app}$ mostly represents the direction between jet axis and LS, we take the viewing angle to be $\varphi=11.8^{\circ}$ to derive distance.

The most natural way to explain the time lag between optical and radio is that the optical emission region is  upstream and the radio emission region is downstream of the jet.
For the most significant lag times, the distance between the emission region of the optical $V$ band and that of 15 GHz, can be expressed by
\beq
\Delta r_{\rm V-15GHz}^{3\sigma} =\frac{\Delta r_{\rm proj}}{\sin \varphi}= \frac{\beta_{\rm app}c \Delta t^{\rm obs}_{\nu_1 \nu_2}}{r_1 (1+z)\sin \varphi} = 4.26^{+0.83}_{-0.79}.
\eeq
where the lag time is taken to be $t_{c}$, and errors considers both $\beta_{\rm app}$ and $\Delta t^{\rm obs}$. For the secondary peaks, the distance is derived as $\Delta r_{\rm V-15GHz}^{2\sigma}=1.26^{+0.41}_{-0.33}{\rm pc}$.
The 15 GHz emission region relative to the base of jet is not given in references. To estimate this scale, VLBI images at two different frequencies are necessary. Assuming the jet is conical, the distance from the 15 GHz core region to the base of jet can be estimated by \cite{Hirotani:2005}
\beq
r_{\rm core}= \frac{r_{\perp}}{\theta}=\frac{0.5\theta_{\rm d}d_{L}}{(1+z)^2\theta},
\eeq
where $r_{\perp}$ is the transverse size of jet at the core position, and $\theta_{d}$ is the core size at certain frequency. One can crudely estimate $\theta_d \approx 0.6 {\rm mas}$ from images of MOJAVE \cite{Lister:2019}. The jet opening angle can be estimated by  $\theta\approx 0.26\Gamma^{-1}=0.026$ \cite{Pushkarev:2009}. The derived 15 GHz core position is located at
\beq
 r_{15\rm GHz, core}\approx 30
\eeq
 away from jet base. A large uncertainty exists due to the
uncertainties of the jet opening angle and core size estimation. If $r_{\rm core}\approx 30 $ pc is true, the location of optical and $\gamma$-ray regions is about $26$ pc away from the jet base. Since
 the typical size of the  BLR region is less than $2{\rm pc}$, so the optical emission region of PMN J$2345-1555$ is most probably beyond the BLR region. \citet{Ghisellini:2013} have presented that the fluxes of the infrared, ultraviolet (UV) and $\gamma-$ray band changed quasi-simultaneously during one flare period, which indicated these emissions are located at one position of the jet. The SED showed that the peak frequency moves toward the bluer part when the flare occurs, and such behavior is interpreted as the emission region moves from within the BLR to the outside. However, it is also possible that the shift of peak frequency is caused by the Doppler boosting, which has no strong constrains on the emission regions.

From the time lag analysis, the $\gamma$-ray emission region coincides with optical. From the LCCF between $\gamma$-ray and radio, it is also possible the $\gamma$-ray emission region is more closer to the 15 GHz core than the optical. By the same procedure,  the distance between $\gamma$-ray emission region and $15$GHz is derived as
$\Delta r_{\rm \gamma-15GHz}^{2\sigma}=1.99^{+0.54}_{-0.46} {\rm pc}$.
 \citet{Hodgson:2017} used the global VLBI network to monitor the OJ 287 for a long period. They found that the $\gamma$-ray emission region is located at downstream of jet coincident with the  radio core region. The flare behavior of the $\gamma$-ray is strongly correlated with the radio flare and the variable structure in the radio images. \citet{Algaba:2018} made the multi-wavelength correlation analysis on 4C 38.41. The $\gamma$-ray  leads to the radio about 70 days, and they found the flare of the $\gamma-$ray is correlated to the radio knot behavior. \citet{Kang:2014} used the SED fitting technique to study the $\gamma$-ray emission region for a sample of bright blazars, and concluded that the seed photons most probably comes from the dust torus. For PMN J2345-1555, the $\gamma$-ray emission region is most probably beyond the BLR if $r_{\rm core}=30$ pc estimation is trustable.

According to Equation (\ref{eq:omegat}), the core-position offset can be estimated by the lag, i.e., $\Delta t^{3\sigma}_{\rm V-15GHz}= 221.81^{+6.26}_{-6.72}$ days. Then, one obtains that
\beq \label{eq:omega3sigma}
\Omega^{3\sigma}_{\rm V-15GHz}=13.07^{+2.54}_{-2.12} {\rm pc \cdot GHz}.
\eeq
If we take $\Delta t^{2\sigma}_{\rm V-15GHz}= -65.80^{+8.67}_{-9.40}$ days, then obtain $\Omega^{2\sigma}_{\rm V-15GHz}=3.89^{+1.23}_{-1.10}$ pc$\cdot$GHz.
The jet opening angle can be estimated by  $\theta\approx 0.26\Gamma^{-1}$, with $\Gamma_{\rm min}=\sqrt{1+\beta_{\rm app}^2}$ \cite{Pushkarev:2009}. Then,the magnetic field at 1 pc can be expressed as is \cite{Pushkarev:2012}
\beq
B_1 \approx 0.042 \Omega_{r\nu}^{3/4} (1+z)^{1/2}(1+\beta_{\rm app})^{1/8}\approx 0.61 {\rm Gauss}.
\eeq
where $\Omega^{3\sigma}$ in Equation(\ref{eq:omega3sigma}) is used. If one considers $\Omega^{2\sigma}$, then obtains $B_1 \approx 0.24$ Gauss. Then, the particle number density is calculated according to Equation (\ref{eq:b1n1}), i.e.,
\beq
N_1=\frac{1533}{ \gamma_{\rm min}} {\rm cm}^{-3}.
\eeq
%
 In the optical emission region, the magnetic field and the electron number density are calculated to be
  $B\approx 0.023\,\,\, {\rm Gauss}$ and $N^* \approx 2.26 /\gamma_{\rm min} \,\,\,{\rm cm^{-3}}$, respectively. For synchrotron radiation, one can estimate the observed peak flux from the formula as follow
\beq
 F_{\nu,\rm max}=(1+z)\delta^3 \left( \frac{N^*_T P_{\nu, \rm max}}{4\pi d_L^2} \right),
\eeq
where $P_{\nu, \rm max}=m_e c^2 \sigma_T \Gamma B /3 q_e$ is the synchrotron peak spectral power, and $N^*_T=\frac{4\pi}{3}R^3 N^*$ is the total lepton number \cite{Sari:1998}. $R=c\delta t_{\rm var}$ can be estimated by the self-correlation LCCF of the optical light curves, where $\delta t_{\rm var}= 48\pm2$ days is the lag where the value of LCCF changes sign. With $\delta=5$ and $z=0.621$, one obtains that $F_{\nu}\approx 2.8 \gamma_{\rm min}^{-1}$ Jansky  for optical $R$ band. The observed flux of $R$ band can be as high as $15$ mJy, see Figure \ref{Fig:7LC}. If  $\gamma_{\rm min}=200$, the estimated value agrees with observation naturally. This indicates that the obtained magnetic fields and particle densities are reasonable parameters.



\subsection{Variable mechanism}

\begin{figure}
\plottwo{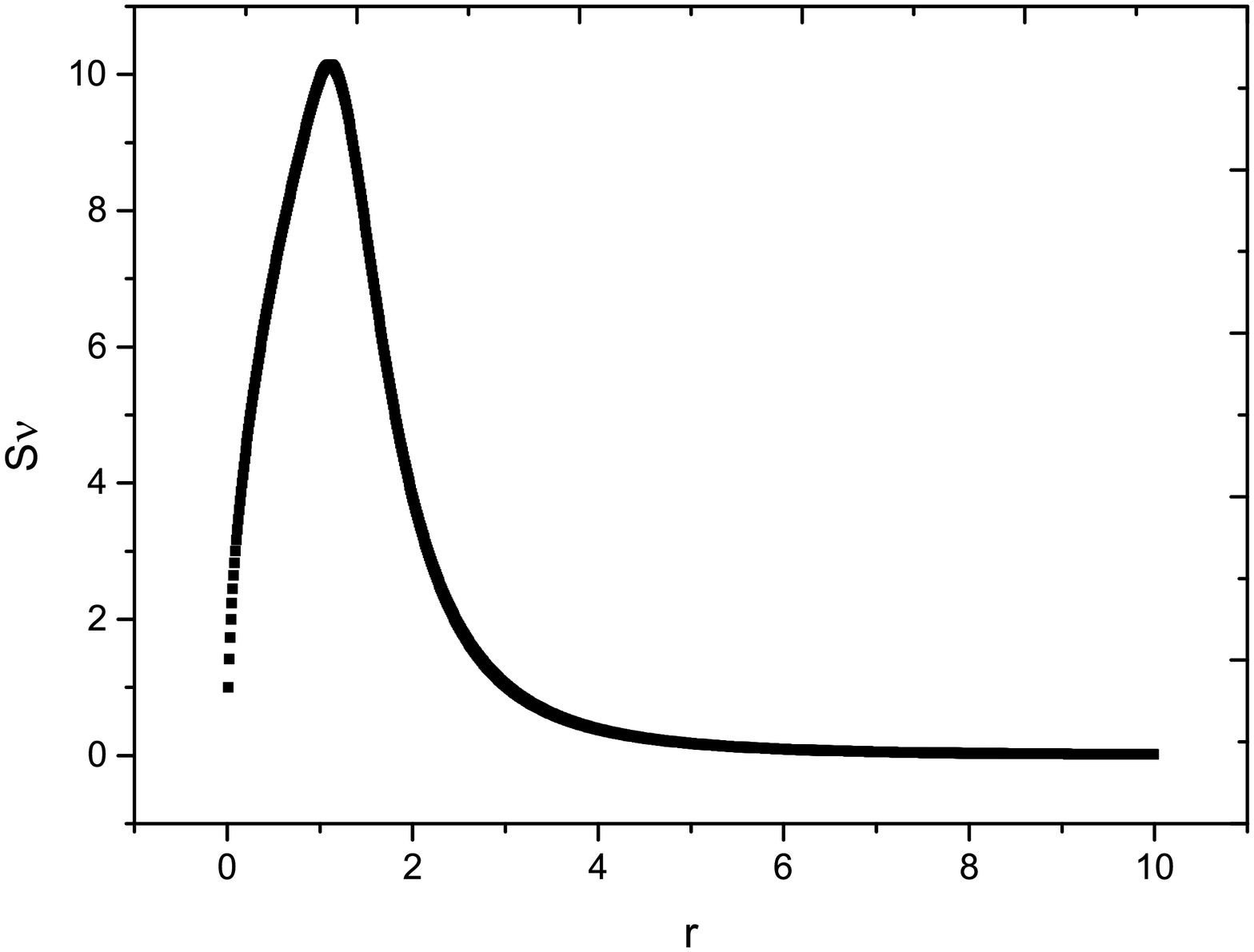}{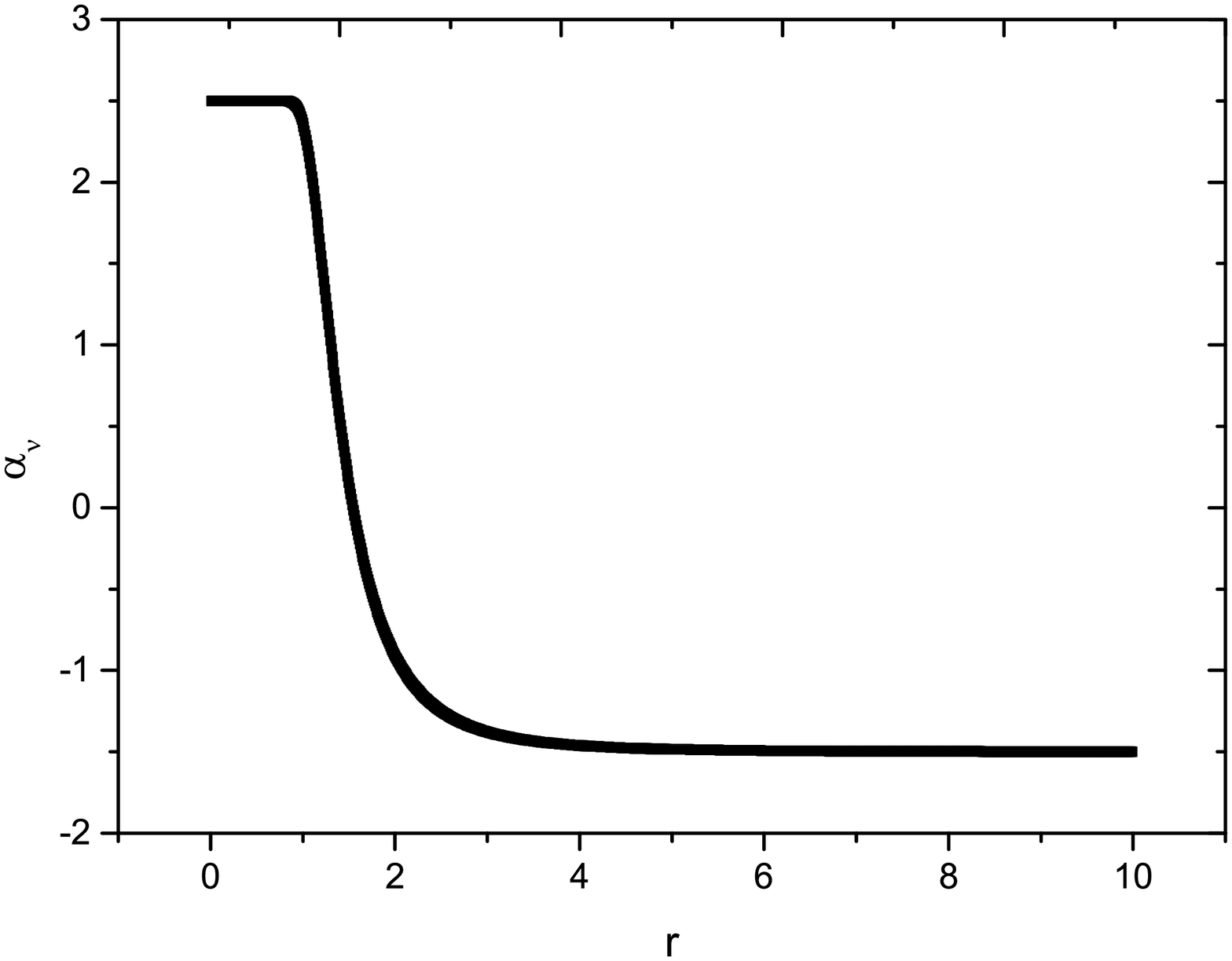}
\caption{The flux $S_{\nu}$ and the spectral index $\alpha_{\nu}$ as functions of $r$ are plotted numerically in the left and right panel, respectively. The units of $S_{\nu}$ is taken arbitrarily.  }
\label{Fig:asn}
\end{figure}

The photosphere emission model can also help us to study the spectral lag phenomenon.
Combing with the scaling law, the flux density is given by
\beq \label{eq:fl1}
S_{\nu}=\Xi(\alpha)\left(\frac{\theta_d}{\rm rad}\right)^2 \left(\frac{\delta}{1+z}\right)^{1/2} B_1^{-1/2} r^{m/2} \nu^{5/2}(1-e^{-\langle\tau_{\nu}\rangle})
\eeq
where $\Xi(\alpha) \equiv \frac{\pi}{4}(\frac{3}{2})^{-\alpha} \frac{e}{c}\frac{ a(\alpha)}{ C(\alpha)} ( \frac{e}{2\pi m_e c})^{-3/2}$.
The spectral index at frequency $\nu$ can be derived by
\beq \label{eq:alphanu}
\alpha_{\nu}=\frac{d \log S_{\nu}}{ d \log \nu}=\frac{5}{2}+(\alpha-\frac{5}{2}) \frac{\langle \tau_{\nu}\rangle e^{-\langle \tau_{\nu}\rangle}}{1-e^{-\langle \tau_{\nu}\rangle}}.
\eeq
where $\alpha$ is the spectral index of the synchrotron radiation at the optically thin region.
In order to discuss how the opacity and flux varies along the jet, we set $m=1,n=2$ and $\alpha=-1.5$. In the upstream jet, the optical depth can be much larger than one. One has $S_{\nu} \propto r^{1/2} \nu^{5/2}$. However, the optical depth decreases dramatically along the jet, i.e., $\langle\tau_{\nu}\rangle \propto r^{-4}$. The photosphere occurs for $\langle\tau \rangle=1$, one can derive that  $\alpha_{\nu} \approx 1+0.58 \alpha$. For  $\alpha=-1.5$, one has  $\alpha_{\nu}\approx 0.13$, which roughly agrees with the observed flat spectral index in the core region \cite{Potter:2012}. When the optical depth becomes much less than one, one has $S_{\nu} \propto r^{-7/2}\nu^{-3/2}$.  Equation (\ref{eq:fl1}) reflects that the optical depth plays a key role in curvature effects of the spectrum. From Equation (\ref{eq:fl1}) and (\ref{eq:alphanu}), it is evident that both $S_{\nu}$ and $\alpha_{\nu}$ strongly depends on $\langle \tau \rangle$. The variation of $\langle \tau \rangle$ of the emission region will leads to the variation of both $S_{\nu}$ and $\alpha_{\nu}$, which can explain the observed $3\sigma$ correlation between $\delta V-\delta R$ and the radio light curve. Since $\langle \tau \rangle \propto r^{1-n-m\epsilon}$,  the analysis here presents us an $r$-dependent model of $\alpha_{\nu}$ and $S_{\nu}$. In Figure(\ref{Fig:asn}), their dependence on $r$ are plotted without considering physical parameters and units. The flux $S_{\nu}$ is plotted according to the formula $S_{\nu}=10r^{1/2}(1-e^{-5r^{-4}})$.  The plots of  $S_{\nu}$ and $\alpha_{\nu}$  agree well with both the flux contours and spectral index maps in VLBI radio images, see Figure 1 and 6 in \cite{OSullivan:2009}.

The $r$-dependent spectral index model also provides us a new  mechanism to explain both  the bluer-when-brighter in the flare state and redder-when-brighter in the quiescent state. The upstream and the downstream emission regions can be considered as two components contributed to the observed total flux. When an emission blob propagates along the jet, the upstream  will become bright first. This component will contribute a relative bluer spectral index to the total flux, which produces a bluer-when-brighter trend.  When the emission blob occurs in the down stream, its kinetic energy has been dissipated previously in the up stream, and the transferred radiative energy is limited, which corresponds to a quiescent state. The spectral index tends to be smaller in the optically thin region.  Besides, it can also accounts for the observed time lags at multi-frequencies. An animation made by 15 GHz VLBI images is presented by \citet{Lister:2016}. This target has a bright core, and the extended knot structures are evident for short time periods. Due to the strong correlation, one can infer that the optical image is very similar to that of the radio. This can explain the observed bluer-when-brighter trend of $R-J$  and the doubtful redder-when-brighter trend of $\delta V- \delta R$  for our target.

The other three kinds of models for the variation of color index includes the contamination from disk or other regions, shock in jet  and Doppler boosting. \citet{Villata:2006} found a redder-when-brighter trend during flares for 3C 454.3, which is explained by the presence of a luminous accretion disk, and the accretion process contributes significantly in the bluer band. The increase of the synchrotron radiation will enhance the flux in the redder band.  Such trend has also been found for the target PKS 1502+106 \cite{Ikejiri:2011}. Also, for 3C 454.3, a color index saturation phenomena was also reported by \citet{Tachibana:2015}. Both the accretion disk model and the  $r$-dependent spectral index model can explain the saturation, since there is a limit for the spectral index of the synchrotron emission. The disk contamination model is also of a two component model.

 The bluer-when-brighter trend can be explained by the particle acceleration in relativistic jet \cite{Kirk:1998}. \citet{Kirk:1998} presented that the spectrum becomes harder when the flux increases, under the assumption that the acceleration time scale of electron is shorter than that the cooling time scale. The bluer band is radiated by the high energy electrons quickly accumulated in a short time. Thus, the spectra becomes hard when the flux rises, and become soft when the flux decays. This produces the bluer-when-brighter trend. A spectral lag relative to the flux will be observed in the same time.  When the acceleration and cooling time scales are almost the same, one observes that the spectra soften when the flux rises, and harden when the flux decays. A redder-when-brighter trend is produced. Thus, the shock in jet model can also produce two inverse trends of variation, which does not conflict with the $r$-dependent spectral index model.

As stated in Section 3, the Doppler boosting in helical jet can also have impact on the spectral index behavior \cite{Raiteri:2017,Sobacchi:2017}. \citet{Raiteri:2017} reported that the long term flux and spectral index variability of CTA 102 is due to curved jet changing in orientation. The spectral index first shows a RWB trend and turns to BWB in flare states. \citet{Sobacchi:2017} indicates that the peak of synchrotron scales as $\nu F_{\nu} \sim \nu^{\xi}$ ($\xi$ can be 1) when the viewing angle is the primary variable parameter.  The decreasing of viewing angle will lead to the enhancement of Doppler factor, and hence the peak frequency and the flux in observer frame. The SED of synchrotron can be fitted by the log-parabolic law \cite{Massaro:2004}, which roughly agrees with spectra in Equation (\ref{eq:fl1}). Supposing that the peak frequency of SED shifts from far-infrared to optical, and then from optical to Ultra-violet, the spectral index of the observing optical bands will change from negative to nearly zero, and from nearly zero to positive, respectively. Correspondingly,  the CI will show a RWB trend first, and turns to BWB trend later. This mechanism can naturally explains the CI behaviors  and spectra of PMN J2345. Thus, the geometrical structure model is another hopeful mechanism.

%

\section{Conclusion}

In this work, we performed the LCCF analysis to study correlations between 8 years light curves of the $\gamma$-ray, optical bands and 15GHz radio. The obtained time delays enable us to study  locations of the $\gamma$-ray and optical emission regions.
The time-resolved spectra analysis is performed to investigate the spectral behaviors at different wavelengths. We present a unified theoretical frame to explain the observed light curve and spectral behavior.
The principle conclusions are given in the follow
\begin{itemize}
  \item Based on the more than $3 \sigma$ significance signals, the optical $V$ band, $R$ band and the infrared $J$ band leads to the 15 GHz radio with  $-221.81^{+6.26}_{-6.72}$, $-201.38^{+6.42}_{-6.02}$  and $-192.27^{+8.26}_{-7.37}$ days, respectively. Combined with the apparent velocity from the VLBI results, the $V$ band emission region locates at $4.26^{+0.83}_{-0.79}$ pc away from core position of 15 GHz.  A  $2\sigma$ correlation peak  indicates that the optical emission region is located $1.26^{+0.41}_{-0.33}{\rm pc}$ pc in the up stream of the $15$ GHz radio core region. $\gamma$-ray probably has the same emission region with optical. A crude estimation gives that the 15 GHz core is about 30 pc away from the jet base, which indicates that the optical and $\gamma$-ray emission regions are beyond the BLR.

  \item We proposed to calculate the core-shift offsets, the magnetic field and lepton number density via time lags among different wavelengths. This method is an alternative for the core-shift measurements, can overcome the low spatial resolution difficulty in optical and $\gamma$-ray observations by using time series. The derived  magnetic fields and lepton number density in the optical emission regions are 0.023 Gauss and $2.26/\gamma_{\rm min} $ cm$^{-3}$, respectively. These derived parameters consistently reproduces the observed flux of optical band for our target.

  \item The  variation of $R-J$ indicates a bluer-when-brighter trend, while  $\delta V-\delta R$ shows complex trend due to the sparse sample. We proposed the $r$-dependent spectral index model, i.e., emissions at different locations in the jet will contribute different spectral indices, to explain the spectral behaviors.  This model is complementary for the shock in jet model, and manifests the opacity effects in the
      radiation processes. The contribution from accretion disk and curvature effects can also leads to the CI behaviors of the target. A comprehensive study is necessary to reveal the variable mechanisms of blazars.
\end{itemize}
In the last, PMN J2345-1555 has a  very compact jet, which is an interesting target for AGN jet property studies via multiple frequency observations.

\acknowledgments

This work has been funded by the National Natural Science Foundation of China under Grant No. U1531105,11403015 and 11873035, the Natural Science Foundation of Shandong Province under Grant No. ZR2014AQ007, ZR2017PA009 and JQ201702,
also partly  by Young Scholars Program of Shandong University, Weihai (No. 20820162003).
Data from the Steward Observatory spectropolarimetric monitoring project were used. This program is supported by Fermi Guest Investigator grants NNX08AW56G, NNX09AU10G, NNX12AO93G, and NNX15AU81G.
This paper has made use of up-to-date SMARTS optical/near-infrared light curves that are available at www.astro.yale.edu/smarts/glast/home.php
This research has made use of data from the OVRO 40-m monitoring program (Richards, J. L. et al. 2011, ApJS, 194, 29) which is supported in part by NASA grants NNX08AW31G, NNX11A043G, and NNX14AQ89G and NSF grants AST-0808050 and AST-1109911.



\vspace{5mm}
\facilities{Fermi(LAT), Steward, SMARTS, OVRO:40m}

%

%

%


%


\begin{thebibliography}{}
\bibitem[Fermi LAT collaboration (2015)]{Fermi:2015}Acero, F., et al., 2015, \apjs, 218, 23

\bibitem[Achterberg et al.(2001)]{Achterberg:2001}Achterberg, A., et al., 2001, \mnras, 328, 393

\bibitem[Ackermann et al.(2015)]{Ackermann:2015} Ackermann, M., et al., 2015, \apj, 813, L41



\bibitem[Algaba et al.(2018)]{Algaba:2018}Algaba, J. C., et al. 2018,\apj, 852, 30


\bibitem[Atwood et al.(2009)]{Atwood:2009}
Atwood, W. B., Abdo, A. A., Ackermann, M., et al. 2009, ApJ, 697, 1071

\bibitem[Britto et al.(2016)]{Britto:2016}Britto., R.J. et al., 2016, \apj, 830, 162

\bibitem[Bonning et al.(2012)]{Bonning:2012}Bonning, E., et al., 2012, \apj, 756, 13

\bibitem[Carrasco et al. (2010)]{Carrasco:2010}Carrasco, L., Carraminana, A., Recillas, E., et al., 2013, Astro. Telegram 2977

\bibitem[Carrasco et al. (2012)]{Carrasco:2012}Carrasco, L., Recillas, E.,Escobedo, G., Carraminana, A., 2013, Astro. Telegram 4608

\bibitem[Carrasco et al. (2013)]{Carrasco:2013}Carrasco, L., Recillas, E., Mayya, D. Y., Carraminana, A., 2013, Astro. Telegram 4736


\bibitem[Chatterjee et al. (2012)]{Chatterjee:2012}Chatterjee et al., 2012, \apj, 749, 191

\bibitem[Ciprini et al. (2010)]{Ciprini:2010}Ciprini, S. 2010,  Astro. Telegram 2408

\bibitem[Cohen et al.(2007)]{Cohen:2007}Cohen M. H., Lister M. L., Homan D. C., et al., 2007, \apj, 658, 232

\bibitem[Donato et al. (2010)]{Donato:2010}Donato, D., Hauser, M., Wagner, S., Hagen, H.,  Astro. Telegram 2972

\bibitem[Edelson \& Krolick (1988)]{Edelson1988}Edelson, R. A. \& Krolik, J. H. 1988, \apj, 333, 649

\bibitem[Fuhrmann et al. (2014)]{Fuhrmann:2014}Fuhrmann, L., Larsson, S., Chiang, J. et al, 2014, \mnras, 441, 1899

\bibitem[Fuhrmann et al. (2016)]{Fuhrmann:2016}Fuhrmann, L., Angelakis, E., Zensus, J.A. et al., 2016, A\&A, 596, 45

\bibitem[Griffith et al. (1994)]{Griffith:1994}Griffith, et al., 1994, \apjs, 90, 179

\bibitem[Ghisellini et al. (2013)]{Ghisellini:2013}{Ghisellini}, G. and {Tavecchio}, F. and {Foschini}, L., Bonnoli, G. {Tagliaferri}, G., 2013, MNRAS, 432, L66-70


\bibitem[Guo et al. (2015)]{Guo:2015}Guo, F., Liu, Y. H., Daughton, W. and Li, H., 2015, \apj, 806, 167

\bibitem[Healey et al.(2007)]{Healey:2007}Healey S. E., Romani, R.W., Taylor, G.B., et al. 2007, \apjs, 171, 61

\bibitem[Healey et al.(2008)]{Healey:2008}Healey et al. 2008, ApJS, 175, 97


\bibitem[Hirotani (2005)]{Hirotani:2005}Hirotani, K., 2005, \apj, 619, 73

\bibitem[Hovatta et al.(2009)]{Hovatta:2009}Hovatta T., Valtaoja E., Tornikoski M., L\"{a}hteenm\"{a}ki A., 2009, A\&A, 494,527


\bibitem[Hodgson et al.(2017)]{Hodgson:2017}Hodgson, J.A., Krichbaum, T. P., Marscher, A.P., et al., 2017, A\&A, 597, 80



\bibitem[Ikejiri et al. (2011)]{Ikejiri:2011}Ikejiri, Y., Uemura, M., Sasada, M., et al. 2011, PASJ, 63, 639

\bibitem[Jackson \& Brown (1991)]{Jackson:1991} Jackson, N. \& Browne, I. W. A., 1991, \mnras, 250, 414-421


\bibitem[Kang et al. (2014)]{Kang:2014}Kang, S. J., Chen, L., \& Wu Q. W., 2014, \apjs, 215, 5

\bibitem[Kaufmann et al. (2010)]{Kaufmann:2010}Kaufmann, S., D'Ammando, F., Gelbord, J., 2010,  Astro. Telegram 2986

\bibitem[Kiehlmann et al.(2016)]{Kiehlmann:2016}Kiehlmann, S., Savolainen, T., Jorstad, S.G., et al.,2016, A\&A, 590, A10

\bibitem[Kirk et al.(1998)]{Kirk:1998}Kirk, J.G., Rieger, F.M., and Mastichiadis A., 1998, A\&A, 333, 452-458

\bibitem[Konigl(1981)]{Konigl:1981}Konigl A., \apj, 1981, 619, 73



\bibitem[Kudryavtseva et al. (2011)]{Kudryavtseva:2011} Kudryavtseva, N. A., Gabuzda, D.C., Aller, M.F., Aller, H.D., 2011, \mnras, 415, 1631

\bibitem[Lasson (2012)]{Lasson:2012}Lasson, S., 2012, arxiv:1207.1459v1

\bibitem[Li et al.(2003)]{Li:2003}Li, W.D. Filippenko, A.V., Chornock, R., Jha, S., 2003, \pasp, 115, 844L


\bibitem[Lico et al.(2014)]{Lico:2014}Lico, R., Giroletti, M., Gomez, J. L., et al., 2014, A\&A, 571, A54

\bibitem[Lister et al. (2009)]{Lister:2009}Lister M. L., et al., 2009, \aj, 137, 3718

\bibitem[Lister et al.(2016)]{Lister:2016}Lister M. L., et al., 2016, \apj, 152, 12

\bibitem[Lister et al.(2019)]{Lister:2019}Lister M. L., et al., 2019, submitted to ApJ



\bibitem[Lobanov(1998)]{Lobanov:1998}Lobanov, A.P. 1998, A\&A,  330, 79


\bibitem[Lyutikov et al. (2005)]{Lytikov:2005}Lyutikov, M., Pariev, V.I. and  Gabuzda, D.C., 2005, \mnras, 360, 869

\bibitem[Marscher et al.(2008)]{Marscher:2008}Marscher, A. P., Jorstad, S. G., D’Arcangelo, F. D., et al. 2008, Nature, 452, 966

\bibitem[Massaro et al.(2004)]{Massaro:2004}Massaro, E., Perri, M., Giommi, P. and Nesci, R., 2004, \aap, 413, 489

\bibitem[Mathew et al. (2010)]{Mathew:2010}Mathew, B., Vadawale, S. V., Naik, S., et al., 2010,  Astro. Telegram 3004

\bibitem[Max-Moerbeck et al.(2014a)]{Max:2014a} Max-Moerbeck W. et al., 2014a, \mnras, 445, 428-436

\bibitem[Max-Moerbeck et al.(2014b)]{Max:2014b} Max-Moerbeck W. et al., 2014b, \mnras, 445, 437-459

\bibitem[O'Sullivan \& Gabuzda(2009)]{OSullivan:2009}O'Sullivan S.P., \& Gabuzda D.C., 2009, \mnras, 400, 26-42

\bibitem[Perterson et al. (1998)]{Peterson:1998}Peterson, B. M., et al., 1998, \pasp, 110, 660-670

\bibitem[Petrov et al. (2006)]{Petrov:2006}Petrov, L., Kovalev, Y.Y., Fomalont, E.B., \& Gordon, D., 2006, \apj, 131, 1872

\bibitem[Potter \& Cotter(2012)]{Potter:2012}Potter, W. J. \& Cotter G., 2012, \mnras, 423, 756-765

\bibitem[Pushkarev et al. (2009)]{Pushkarev:2009}Pushkarev, A.B., Kovalev, Y.Y.,Lister, M. L., and Savolainen, 2009, A\&A, L33-L36

\bibitem[Pushkarev et al. (2010)]{Pushkarev:2010}Pushkarev, A.B., Kovalev, Y.Y., and  Lister, M. L., 2010, \apj, 722, L7-L11

\bibitem[Pushkarev et al. (2012)]{Pushkarev:2012}Pushkarev, A. B. et al., 2012, A\&A, 545, 113

\bibitem[Raiteri et al. (2017)]{Raiteri:2017}Raiteri, C.M., et al., 2017, nature, 552, 374

\bibitem[Richards et al.(2011)]{Richard:2011}Richards, J. L. et al, 2011, \apjs, 194, 29


\bibitem[Schmidt et al.(1992)]{Schmidt:1992}Schmidt, G.D., Stockman, H.S., \& Smith, P.S. 1992, \apj, 398, L57


\bibitem[Sari et al. (1998)]{Sari:1998}Sari, R., Piran, T., Narayan, R., 1998, \apj, 497, L17-L20



\bibitem[Schlafly et al. (1982)]{Schlafly:1982}Schlafly, E. F., 2011, \apj, 737, 103





\bibitem[Shaw et al (2009)]{Shaw:2009} Shaw, M.S., Romani R. W., Healey, S.E., et al., 2009, \apj, 704, 477

\bibitem[Shen et al.(2011)]{Shen:2011}Shen et al., 2011, \apj, 194, 45

\bibitem[Smith et al.(2009)]{Smith:2009}Smith, P.S., et al., 2009, arXiv:0912.3621, 2009 Fermi Symposium, eConf Proceedings C091122.

\bibitem[Sobacchi et al. (2017)]{Sobacchi:2017}Sobacchi, E., Sormani, M. C. and Stamerra, A., 2017, \mnras, 465, 161







\bibitem[Tachibana and Kawai (2015)]{Tachibana:2015}Tachibana, Y., and Kawai, N. 2015, arXiv:1502.03610v1

\bibitem[Timmer and Koenig (1995)]{Timmer:1995}Timmer, J., \& Koenig M., 1995, A\&A, 300, 707




\bibitem[Urry and Padovani (1995)]{Urry:1995}Urry, C.M., \& Padovani P., 1995, PASP, 107, 715

\bibitem[Villata et al.(2006)]{Villata:2006}Villata, M., et al., 2006, A\&A, 453, 817-822

\bibitem[Welsh (1999)]{Welsh:1999}Welsh, W. F., 1999, PASP, 111, 1347

\bibitem[Wu et al.(2018)]{Wu:2018}Wu, L. H., Wu Q. W., Yan, D. H., Chen, L. \& Fan, X. L., 2018, \apj, 852, 45

\bibitem[Yan et al.(2018)]{Yan:2018}Yan D.H., Wu, Q.W., Fan, X. L., Wang, J.C. \& Zhang, L., 2018, \apj, 859, 168

\bibitem[Zamaninasab et al.(2014)]{Zamaninasab:2014}Zamanimasab M., et al., 2014, Nature, 510, 126


\end{thebibliography}
\end{document}